\documentclass[preprint]{aastex}
\usepackage{graphicx,natbib}

\begin{document} 

\title{{Deviations from the Local Hubble Flow. I. The Tip of the Red Giant Branch as a Distance Indicator}}
\author{Bryan\ M\'{e}ndez\altaffilmark{1}, Marc\ Davis\altaffilmark{1,2},
        John\ Moustakas\altaffilmark{3}, Jeffrey\ Newman\altaffilmark{1}, 
        Barry F.\ Madore\altaffilmark{4}, and Wendy L.\ Freedman\altaffilmark{4}}
\altaffiltext{1}{Department of Astronomy, 601 Campbell Hall, University of California, Berkeley, CA 94720-3411.}
\altaffiltext{2}{Department of Physics, 366 LeConte Hall, University of California, Berkeley, CA 94720-7300.}
\altaffiltext{3}{Steward Observatory, 933 North Cherry Avenue, University of Arizona, Tucson, AZ 85721}
\altaffiltext{4}{Observatories of the Carnegie Institute of Washington,
                 813 Santa Barbara St., Pasadena, CA 91101.}

\begin{abstract}
The properties of the velocity field in the local volume ($cz < 550$ km s$^{-1}$) have been difficult to constrain due to a lack of a consistent set of galaxy distances. The sparse observations available to date suggest a remarkably quiet flow, with little deviation from a pure Hubble law. However, velocity field models based on the distribution of galaxies in the 1.2 Jy IRAS redshift survey, predict a quadrupolar flow pattern locally with strong infall at the poles of the local Supergalactic plane. We probe this velocity field and begin to establish a consistent set of galactic distances. We have obtained images of nearby galaxies in the $I$ and $V$ band from the W.M. Keck Observatory and in F814W and F555W filters from the {\em Hubble Space Telescope}. Where these galaxies are well resolved into stars we can use the Tip of the Red Giant Branch (TRGB) as a distance indicator. Using a maximum likelihood analysis to quantitatively measure the $I$ magnitude of the TRGB we determine precise distances to seven nearby galaxies: Leo~I, Sextans~B, NGC~1313, NGC~3109, UGC~03755, UGC~06456, and UGC~07577. 

\end{abstract}


\section{\bf Introduction}

\subsection{\bf Local Deviations from Hubble Flow}

Although tremendous progress has been made in recent years in mapping
the large-scale gravitational field (out to scales of $cz \sim 10,000$
km s$^{-1}$), relatively little data are available to study the velocity
field of galaxies in the local neighborhood ($\leq 500$ km s$^{-1}$).  The
major impediment to constraining the local flow is the lack of a
consistent set of distances to nearby galaxies. Locally, the
deviations from a pure Hubble law have been predicted to be large and
measurable, as much as 1 magnitude in distance modulus.

\begin{figure}
\begin{center}
\includegraphics[height=7.25in,angle=270]{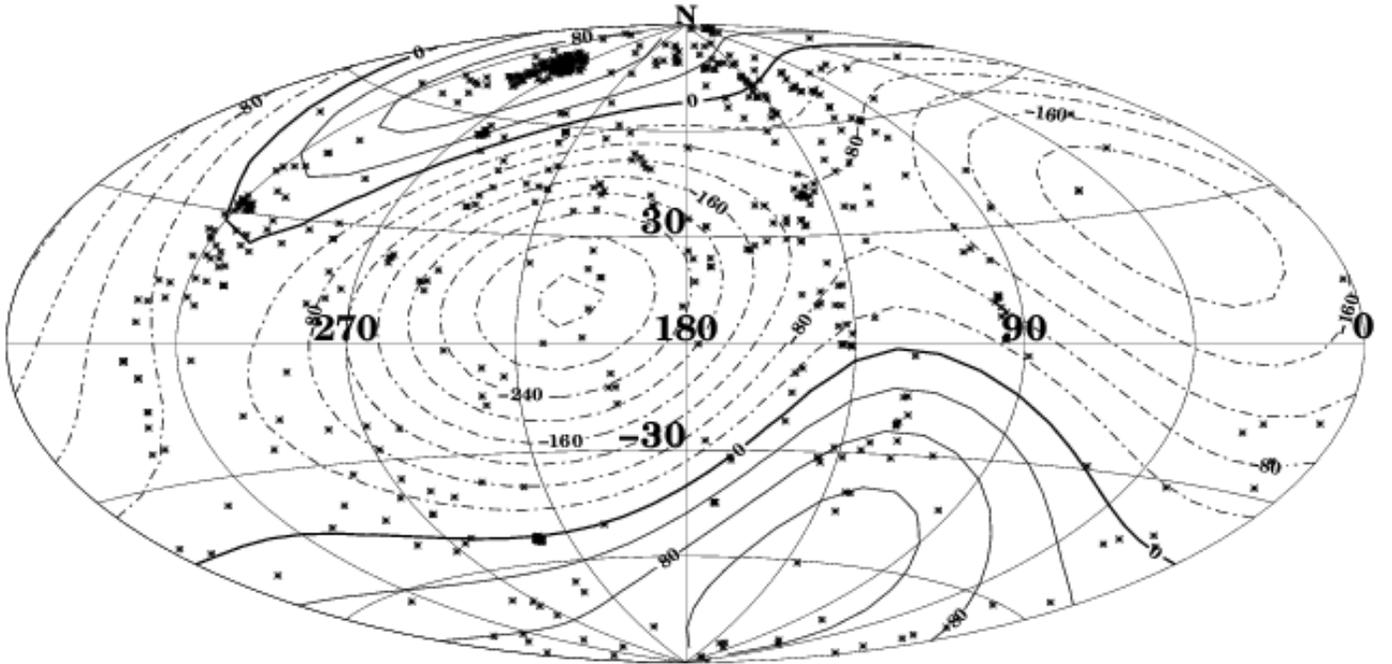}
\end{center}
\caption{Full-sky contour  map  (in Galactic co-ordinates) of the
predicted radial velocity field on a shell at redshift $cz = 500$ km s$^{-1}$
(in the   frame  of  the Local Group)  derived  from the  IRAS
galaxy distribution (see \citet{nusser94}, for details).  Contours  are in units of 50  km s$^{-1}$ with the dash--dot lines representing infall peculiar velocity and solid lines representing outflow peculiar velocity. The amplitude  of
the flow is  almost linear in  $\beta  \equiv \Omega^{0.6}/b_I$,  where
$b_I$ is the linear bias in  the IRAS galaxy  distribution relative to
the mass distribution.  The plot assumes $\beta = 0.6$.
Galaxies with $cz < 550$ km s$^{-1}$ are plotted.}
\label{figure:skydist}
\end{figure}

If gravitational instability has generated  all the
large  scale  structure in the  Universe, then it  must  also   be
responsible  for  the extremely anisotropic  local galaxy distribution
 (see Fig.~\ref{figure:skydist}) \citep{peebles93}. Under this assumption, it is possible to predict the velocity field from this distribution.
The catalogs from such full--sky galaxy  redshift surveys as the 1.2 Jy IRAS survey \citep{fisher94} allow us to estimate the large
scale gravity field out to a redshift $cz \approx 8,000$ km s$^{-1}$.  In the
limit of linear perturbations at late  times, one expects the peculiar
velocity at a point in space to  be  proportional to the gravity field
at that point. 
Fig.~\ref{figure:skydist} demonstrates that according to linear  theory,
the local velocity field is  expected  to  show  substantial deviations   
from uniformity, with quadrupole distortions dominating. 
Outflow is directed
toward Virgo and Fornax on opposite sides of the sky, while substantial
infall occurs from  the poles of the  local  distribution.   Such a  flow is
expected within any gravitational  model and would act to depopulate galaxies from the local poles. The bulk of the local galaxy distribution does reside
on the positive and zero contours of Figure~\ref{figure:skydist} (the local supercluster
plane), while the regions of negative peculiar velocity are deficient
but not entirely devoid of galaxies.  
It is possible that linear theory does not apply on this scale,
but in  that case one would expect  a more chaotic flow of similar amplitude.

Observations of local deviations from Hubble flow do not
really support the expectations of Fig.~\ref{figure:skydist}, especially in the
southern direction (see the review by \citet{burstein90}).  However, the data are
sparse because large spiral galaxies suitable for Tully-Fisher or Cepheid Period--Luminosity 
analysis are very rare locally, and the sky coverage, particularly in
the expected infalling regions, is poor.  We have thus set out to provide
new distances to a complementary sample of nearby galaxies that were chosen to lie near the maximum predicted infall.

\subsection{\bf A Consistent Distance Scale for Nearby Galaxies: The Tip of the Red Giant Branch (TRGB)}

According to  stellar  evolution theory,  the tip of the first--ascent
red--giant  branch  (TRGB) marks  the onset of the core--helium  flash in
low--mass stars.  Observationally, this phenomenon causes a
distinct  and abrupt termination  of the bright  end of the red--giant
branch  luminosity function.   This discontinuity  translates directly
into an excellent distance indicator: the bolometric luminosity of the
TRGB for low--mass  stars is predicted to vary by only  $\sim 0.1$ mag
for ages ranging  from 2  up to 15  Gyr \citep{iben83} and for metallicities encompassing the entire range represented by Galactic globular clusters ($-2.1 < [Fe/H] < -0.7$) \citep{salaris97}. The discontinuity is also found empirically to be stable at the $\sim 0.1$ mag level in the $I$ band for the same set of stellar properties.

The TRGB method has a strong empirical, as well as physical, basis.
Distances obtained using the TRGB agree with the Cepheid Period--Luminosity relation
(where direct comparisons have been made) at a $\pm$5\% 
level, e.g. \citet{lee93,sakai96,udalski01} -- an exception is for M33: \citet{kim02}. Furthermore, since all visible galaxies must have had a first generation of stars, 
the TRGB method is applicable to populations in all morphological
types, including spiral, elliptical, and even irregular
galaxies, provided their metallicity is sufficiently low. The TRGB method also requires far less telescope time for observations than Cepheids since repeat visits are unnecessary, a considerable advantage. A good review of the TRGB, its applications and theoretical underpinnings, is given in \citet{madore98}.

We have used the Tip of the Red Giant Branch to obtain distances to a sample of local galaxies (mostly dwarfs) in order to derive their peculiar velocities.  In \S~2 we  discuss the observations we have made of several galaxies using both the W.M. Keck observatory and the {\em Hubble Space Telescope} ({\em HST}). In \S~3 we present the methods for measuring the TRGB that we have employed and include detailed information on the analysis and error propogation. Finally, in \S~4 we present the Color--Magnitude Diagrams, Luminosity Functions, and derived distances for several of the galaxies studied. We discuss the derived peculiar velocities and their implications in paper II \citep{mendez02b}.

\section{\bf Data}

To select targets for observation, we assembled an all-sky sample of 502 galaxies likely to be in the local volume (cz $_{\sim}\!\!\!\! ^{<} 550$  km s$^{-1}$) . This sample consists of all objects identified as galaxies (Type G) with heliocentric radial velocities less than +900 km s$^{-1}$ within the NASA/IPAC Extragalactic Database (NED)\footnote{The NASA/IPAC Extragalactic Database (NED) is operated by the Jet Propulsion Laboratory, California Institute of Technology, under contract with the National Aeronautics and Space Administration.}. We then transformed all velocities to the Local Group frame \citep{karachentsev96} and selected those having radial velocities less than +550 km s$^{-1}$. Their distribution on the sky (in Galactic coordinates) is shown in Fig.~\ref{figure:skydist}.
Undoubtedly, many ($\sim 100$) of these galaxies are not actually within the local neighborhood but are cluster members whose large peculiar velocities arising from virial motions bring them into our sample. This is clearly the case toward the Virgo cluster ($l = 283.78^{\circ} $, $b = 74.49^{\circ}$).
We are most interested in galaxies located in the regions of strong infall predicted by the flow models of  \citet{nusser94} and \citet{baker98}. We have observed a subset of these galaxies with the Keck Telescope in Hawaii and/or the {\em Hubble Space Telescope}. 

\subsection{\bf Keck Observations and Reduction}

We observed target galaxies on the nights of 1997 December 22 and 23 under photometric conditions using the Low Resolution Imaging Spectrometer (LRIS) \citep{oke95}. At that time LRIS operated at the Cassegrain focus of the 10-m Keck II telescope at the W.M. Keck Obseratory located on Mauna Kea, Hawaii (it now resides on Keck I). The instrument includes a backside-illuminated Tek $2048 \times 2048$ CCD detetector with an imaging scale of 0.215 arcseconds/pixel and a field of view of $6\times8$ arcminutes. We read out the CCD in the two-amplifier mode.

Each galaxy was observed through both Johnson $V$ and Cousins $I$ broadband filters. For the brightest, best resolved galaxies we took one exposure in each filter, of typically 300 sec duration . For the fainter, less well--resolved galaxies we took between three and eight 250--500 second exposures in $I$ and one to two 300--800 second exposures in $V$. See Table~\ref{table:obs} for details. Over both nights four Landolt standard star fields were observed \citep{landolt92} in order to calibrate the photometry.

The images were reduced according to standard procedures using software written in IDL. The CCD frames were subtracted for bias level (determined from the overscan region appropriate to that amplifier) and dark current. The right side of each frame was scaled to match the left using the gain (1.97 $e^{-}$/count for the left amplifier, 2.1 $e^{-}$/count for the right) and the respective bias levels. In $I$ sky flats were constructed using all of a single night's image frames themselves, while in $V$ dome flats were used to remove pixel-to-pixel variations. The resulting image frames were found to be uniform in sensitivity across their widths to better than 1\% .
Frames for individual galaxies were registered and combined through a median filter to remove cosmic rays and increase the signal-to-noise ratio. These combined frames were used for finding stars.

 Due to
geometric and optical distortions, the point spread function (PSF) varies as a function of
position on the LRIS CCD chip.  It was therefore necessary to compute
an empirical PSF for each image from a hand-selected list of $10-30$
bright, unsaturated stars that had no stellar neighbors within
$\sim3\times$FWHM. The PSF for each data image was modeled using the DAOPHOT package within IRAF\footnote{IRAF is
distributed by the National Optical Astronomy Observatories, which is
operated by the Association of Universities for Research in Astronomy,
Inc., under contract to the National Science Foundation.}.  Stars were identified and instrumental magnitudes
were computed using the FORTRAN versions of DAOPHOT II/ALLFRAME
\citep{stetson87,stetson94}.  These programs, modified for FITS I/O,
were obtained from K. Gebhardt at the University of California, Santa
Cruz.  
Several of the fainter targets were not well resolved into stars, and thus were not usable for determining TRGB distances. Thus, we did not perform full photometry on these objects.

The instrumental magnitudes $m_{\rm inst}$ and magnitude errors for each
image in a filter were combined using a robust mean (a weighted mean iteratively refined by a factor of $1/(1 + \chi_{i}^2)$ for each weight, $w_i$) and converted
to standard magnitudes $m_{\rm stand}$ with the equation
\begin{equation}
 m_{\rm stand} = m_{\rm inst} + 2.5\log (t) + {\rm ac} - {\rm c_{1}} - {\rm c_{2}X} - {\rm c_{3}}(V-I) - {\rm c_{4}X}(V-I)  
\end{equation}
where $t$ is the exposure time, ac is the aperture correction, X is the airmass for each observation, and c$_{1}$, c$_{2}$, c$_{3}$, and c$_{4}$ are constants. The aperture correction, ac, was computed by comparing the magnitudes of bright isolated stars in each object found through an aperture of 40 pixels (8.6 arcsec) in diameter to the magnitude found via PSF fitting. Values of the aperture corrections range from $0.00$ to $-0.10$ mag, with standard errors of $0.01$ to $0.03$ mag. The broadband filters on LRIS are very close to the standard system, so the color terms, c$_{3}$ and c$_{4}$, were found to be negligable. The extinction coefficients (in magnitudes/airmass) given by c$_{2}$ are 0.12 at 550 nm and 0.07 at 800 nm \citep{beland88}. The magnitude zero point, c$_{1}$, was computed by using the above information and aperture photometry of both Landolt standard stars and the additional stars in Landolt fields recently calibrated to the standard system \citep{stetson00}. The interactive IRAF task {\em fitparams} in the {\bf noao.digiphot.photcal} package was used to compute a non-linear least squares fit of the observed magnitudes to the calibrated magnitudes of these standards. The resulting zero points have an rms error of 0.02 mag in both $I$ and $V$. 

\subsection{\bf {\em HST} Observations and Reduction}

We also obtained images of thirteen galaxies with the Wide Field Planetary Camera 2 (WFPC2) aboard the {\em Hubble Space Telescope}\footnote{Observations made with the NASA/ESA {\em Hubble Space Telescope}, obtained at the Space Telescope Science Institute, which is operated by the Association of Universities for Research in Astronomy, Inc., under NASA contract NAS 5-26555. These observations are associated with proposal \#~8199.} between August 13, 1999 and May 9, 2000 (program 8199, PI M. Davis). Galaxies were observed in two exposures from 1200 - 1300s each in the F814W filter; three objects were also observed with two 1200 - 1300s exposures in the F555W filter. Table~\ref{table:obs} details the WFPC2 observations. These thirteen galaxies were chosen specifically because they are located in regions of the sky where strong inflow is predicted by the IRAS data, especially around $l = 170^{\circ} - 240^{\circ}$ and $b = 10^{\circ} - 30^{\circ}$.

The calibrated and data--quality images were retrieved from the Space
Telescope Science Institute (STScI) where preliminary CCD processing,
including bias- and dark-subtraction, flat--fielding, and a shutter--shading 
correction, are automatically applied using calibration data
closest to the time of the observation.  Details of these initial
reductions are given in \citet{burrows94} and
\citet{holtzman95}.

For photometric measurements we used the PSF-fitting package
HSTphot\footnote{The source code and documentation for HSTphot can be
obtained at http://www.noao.edu/staff/dolphin/hstphot.} developed by
A. Dolphin specifically for analyzing HST/WFPC2 data.  HSTphot has
been used recently in a number of stellar population studies
\citep{wyder01,saha01,dolphin01,seitzer01}.  It has been optimized for
the undersampled conditions present in WFPC2 data and works very well
in crowded fields.  HSTphot is written in C and runs non-interactively from
the command line.  In our
reductions, HSTphot was an order of magnitude faster than
ALLFRAME, and identified $10-30\%$ more stars.  The algorithms
used by HSTphot most resemble those of DoPHOT \citep{schechter93};
tests of HSTphot photometry including a comparison with an independent DoPHOT
analysis are discussed in \citet{dolphin00a}.  Our own limited tests of
HSTphot reliability as measured against ALLFRAME photometry showed no
gross systematics, even at faint magnitudes.  

HSTphot determines stellar magnitudes using a library of Tiny Tim PSFs
\citep{krist96} that have been built for all the commonly used WFPC2
filters.  The PSFs are
modified internally for geometric distortions using the equations of
\citet{holtzman95}, and corrected for the 34$^{\rm th}$ row error
\citep{shaklan95}.  The final stellar magnitudes are calibrated using
the charge-transfer efficiency and zero-point corrections derived in
\citet{dolphin00b}.

The data were analyzed using HSTphot according to the following
recipe: (1) mask bad pixels and columns using the data-quality image;
(2) reject cosmic rays; (3) flag and mask hot pixels; (4) generate a
sky image; (5) find stars and make photometric measurements; and (6)
determine and apply aperture corrections.  The output from the
\emph{multiphot} procedure in HSTphot was trimmed according to the following criteria to
generate the final, calibrated star list: $\chi^2 <5$, $-0.5<\,{\rm
sharpness\,}<0.5$, ${\rm S/N}\,>3.5$, and ${\rm class}\,<3$.

We augmented our sample of HST data by searching the WFPC2 archive\footnote{Data are based on observations made with the NASA/ESA {\em Hubble Space Telescope}, obtained from the Data Archive at the Space Telescope Science Institute, which is operated by the Association of Universities for Research in Astronomy, Inc., under NASA contract NAS 5-26555. These observations are associated with proposal \#~6276.} for suitable nearby galaxies with observations in the F814W and F555W
filters. The blue compact dwarf galaxy UGC~06456 was found to have a well defined TRGB as discussed in more detail in \S~4.3.6. These data were reduced in
the same way as described above.

\section{\bf Measuring the TRGB}

\subsection{\bf Detection of the TRGB}

Early estimates of the apparent magnitude of the TRGB used a simple ``eyeball'' estimate of its location in a Color-Magnitude Diagram (CMD), e.g. \citet{mould86}. In the 1990s more quantitative methodologies were introduced \citep{lee93}. In our analysis we have used two distinct methods to quantitatively estimate the TRGB magnitude from the shape of the luminosity function of a population of stars. 

After the exhaustion of hydrogen fuel in the core of a star there is a rapid increase in the rate at which its luminosity will change with time. Because this evolution is so quick, stars that appear above the main sequence turnoff point in the CMD must have had the same Main Sequence lifetime, and hence the same initial mass, to within a few percent. Therefore, the luminosity function of a star cluster (or single population of stars) past the Main Sequence is primarily a reflection of the rate of luminosity evolution for those stars. See the review by \citet{renzini88} for details in interpreting the luminosity functions of populations of evolved stars. 

 Low-mass, post-main-sequence stars develop an inert isothermal core supported by electron degeneracy pressure. The luminosity of these stars is almost entirely dependent on the mass of that core, because all of their energy generation occurs in a thin hydrogen burning shell surrounding the core. The temperature of the shell is set by the amount of potential energy converted to heat by the gravitational contraction of the core. Once on the red giant branch (the vertical track in the CMD) the stars' rate of luminosity evolution is primarily dependent on the rate of increase of the mass of the core. 

Due to the simple dependence of luminosity on core mass, red giant branch luminosity functions are expected to be very simple and depend only very weakly on parameters such as age, metallicity, or mass of the envelope of the star. Indeed theoretical and empirical studies find the bright end of red giant branch luminosity functions to have a simple power-law behavior; i.e., they are straight lines in the magnitude-log(counts) plane \citep{zoccali00}.
Models all predict essentially the same power-law slope and are in excellent agreement with observations.

The luminosity function of the red giant branch terminates at the TRGB, the core mass at which the helium flash occurs, and the stars then adjust their structures rapidly and move to a new location in the CMD (the Horizontal Branch of helium core--burning stars). The magnitude of the TRGB is the location in the luminosity function where the red giant branch power--law truncates. Brightward of that break are other populations of stars (asymptotic giant branch and red supergiant stars mostly) but they are rare and have a distinctly different luminosity function behavior. 

From our data we have determined that if the red giant branch luminosity function power-law is given as
\begin{equation}
 N(m)dm \propto 10^{a m},  
\end{equation}
 where $m$ is the apparent magnitude, then $a = 0.30 \pm 0.04$. This fixed slope can also be found in numerous other studies of red giant branch luminosity functions \citep[for example]{zoccali00} though that fact is rarely explicitly stated.

\subsubsection{\bf Discontinuity Detection}

The first TRGB estimation method we have applied simply uses the fact that there will be a discontinuity in the luminosity function at the location of the TRGB. The first derivative of the luminosity function should therefore show a large peak at the discontinuity. Early applications often used histograms to approximate the luminosity function and a Sobel edge--detection filter to estimate the first derivative \citep{lee93}, but this had the major drawback that the estimates could be strongly dependent on the histogram binning size. Later \citet[hereafter referred to as SMF96]{sakai96} updated this method by employing a Gaussian-smoothed luminosity function, $\Phi (m)$, and a continuous edge--detection function
\begin{equation}
E(m) = \Phi (m + \bar{\sigma}_{m}) - \Phi (m - \bar{\sigma}_{m}), 
\end{equation}
where $\bar{\sigma}_{m}$ (as defined by SMF96) is the mean photometric error within a bin of $\pm 0.05$ mag about magnitude $m$.

We have used the method of SMF96, with some minor modifications, to estimate the TRGB magnitude of our objects. Due to the natural power--law form of the luminosity function we are most concerned with fractional changes. Thus, we use a logarithmic edge--detection function,
\begin{equation} E(m) = \log_{10}(\Phi (m + \bar{\sigma}_{m})) - \log_{10}(\Phi (m - \bar{\sigma}_{m})).
\end{equation}
Another modification to the SMF96 method is the definition of $\bar{\sigma}_{m}$ used. For each galaxy observed we determine a polynomial fit to our data for the photometric flux errors as a function of flux. The function is then transformed to magnitude error as a function of magnitude. Thus, we define $\bar{\sigma}_{m}$ as simply the photometric error given by the fit at magnitude $m$.
 The region of the luminosity function brighter than the TRGB is sparsely populated (mostly by AGB stars) and thus very noisy; the large first derivative noise spikes can make it difficult to robustly identify the TRGB automatically. We thus weight the edge--detection response by the Poisson noise in the smoothed luminosity function. Even that fails to quiet some noise spikes and we are thus forced to limit this analysis to the region of the luminosity function that we know by inspection contains the TRGB. See the figures in \S~4.3 for results of the edge--detection (e.g. Fig.~\ref{figure:leoi_ccut_lf})

\subsubsection{\bf Maximum Likelihood Analysis}

Binning the data is not necessary, and we have employed a maximum likelihood analysis similar in spirit to the luminosity function
analysis of \citet[see also references therein]{sandage79}.  Define the luminosity distribution function of 
stars in the galaxy under investigation to be
$\phi(m | {\bf a})$  where $m$ is the apparent magnitude and {\bf a} is the 
list of parameters defining the function.  The probability $P$ that star $i$
has an observed magnitude $m_i$ within a range $dm$ is then 
\begin{equation}
P_i = {\phi(m_i | {\bf a}) dm \over \int_{m1}^{m2} \phi(m | {\bf a})dm }  
\end{equation}
where $m1$ and $m2$ are lower and upper bounds for the validity of
the parameterized distribution function.

The likelihood ${\cal L}$ that the observed distribution of stellar magnitudes
is described by this function is simply the product of the individual
probablities, since each star is drawn randomly from this distribution.
  It is most convenient to consider $\ln{\cal L}$ 
\begin{equation}
\ln{\cal L} = \sum_i \ln{P_i} = \sum_i \ln{\phi(m_i | {\bf a})} -
                 N \ln{(\int \phi(m) dm)}, 
\end{equation}
where the sum extends over all stars $i$ with $N$ stars in total.

For this analysis we use a simple, broken power-law model to describe the
luminosity distribution function, but we also take into account the smoothing
induced by measurement errors for the apparent luminosities.  We use the form
\begin{equation} 
\phi(m) =  \int g(m') e(m-m')dm'   
\end{equation} 
and define $g(m)$ as

\begin{equation}
g(m) =  \left\{ \begin{array}{cl}
                10^{a(m-m_{\rm TRGB})}     & m-m_{\rm TRGB} >0 \\
                10^{b(m-m_{\rm TRGB}) - c}  & m-m_{\rm TRGB} < 0 \\
                \end{array}  \right.
\end{equation}

We fix $a = 0.3$ but allow as free parameters the TRGB magnitude $m_{\rm TRGB}$,  
the slope of the luminosity function brighter than the TRGB $b$, and the 
strength of the discontinuity at the TRGB, $c$. 
The smoothing function $e(x)$ is the same error function mentioned in \S~3.1.1 used to estimate $\bar{\sigma}_{m}$. It is a function of the apparent magnitude but not of the parameters of the fit. 

\begin{figure}
\begin{center}
\includegraphics[height=8in]{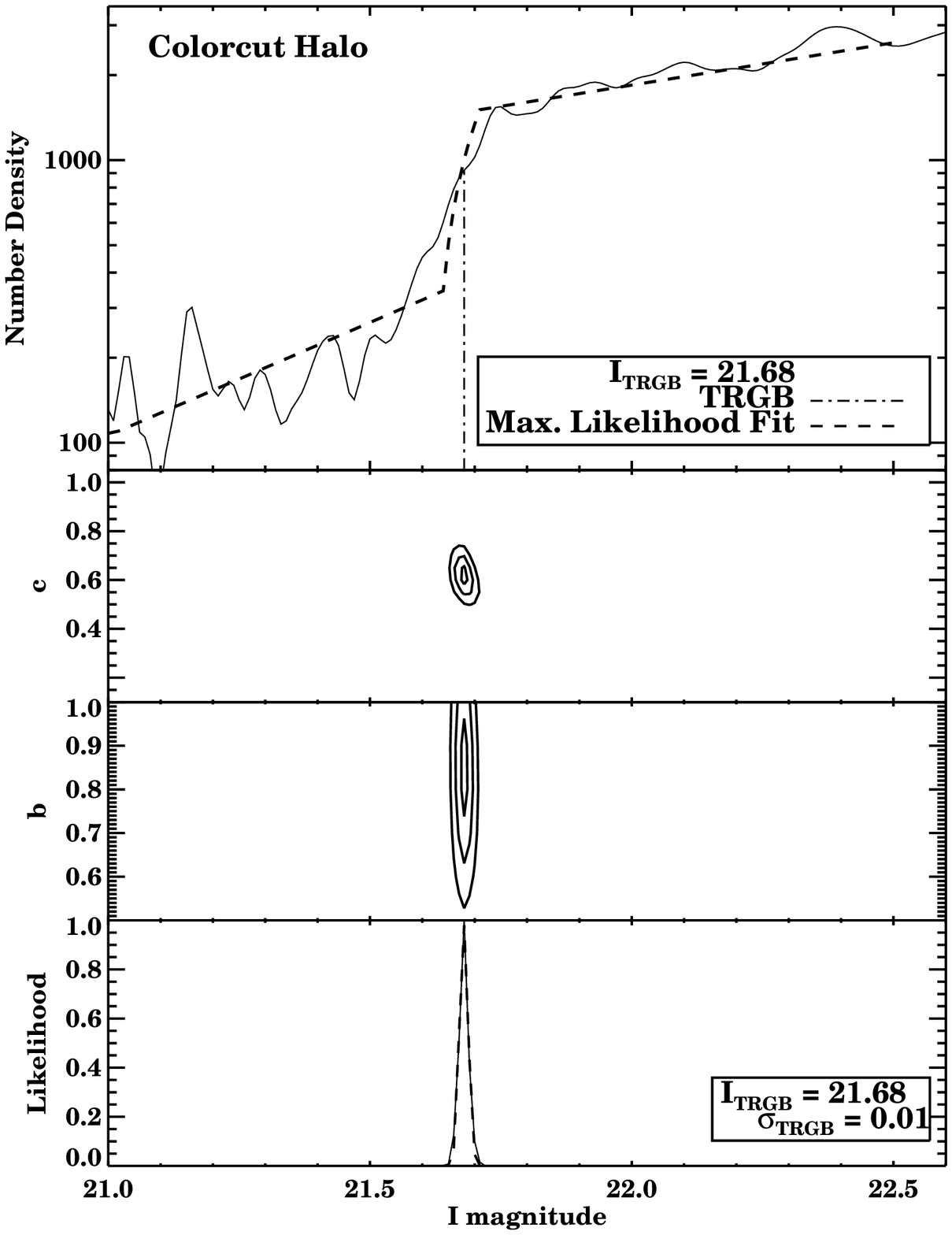}
\end{center}
\caption{Luminosity function of stars in the halo of Sextans B with $0.6 \leq (V-I) \leq 1.6$ mag (top panel). Overplotted is the best--fit model luminosity function determined via a maximum likelihood analysis. The second and third panels plot $1\sigma$, $2\sigma$, and $3\sigma$ contours of the likelihood function parameters $c$ and $b$. The fourth panel plots the marginalized likelihood. A Gaussian fit with $\sigma_{\rm TRGB} = 0.01$ mag is overplotted as a dashed line.}
\label{figure:SextansB_halo_ccut_like}
\end{figure}

Our procedure is then to compute the likelihood function on a grid in 
the parameters $m_{\rm TRGB}$, $b$, and $c$ and to determine the point of maximum
likelihood.  The method used is self--normalizing; the likelihood contours 
are quite peaked, with $3\sigma$ error widths in $m_{\rm TRGB}$ usually less than $0.1$ mag (e.g. Fig~\ref{figure:SextansB_halo_ccut_like}). To test for covariance we marginalize the likelihood by computing
\begin{equation}
{\cal L}(m) = \int_{0.0}^{1.2} \int_{0.5}^{1.0} {\cal L}(m,b,c) db\, dc .
\end{equation}
Covariance with the variables, $b$, and $c$ is found to be negligible, and the peaks in most of the marginalized likelihoods occur at the same magnitude as the peaks in the full likelihoods. In the cases reported in \S~4.3 there are no differences, and the largest difference in all cases is only $0.06$ mag. We fit gaussians to the marginalized likelihoods and find that the $1\sigma$ errors in $m_{\rm TRGB}$ are typically only a few hundredths of a magnitude.

The best fitting $\phi(m)$ curves for each galaxy are displayed in the figures of \S~4.3. Note that these fits are quite acceptable in each case.  
Smoothing by the photometric errors converts the discontinuity in $g(m)$ at $m = m_{\rm TRGB}$ into a continuous distribution $\phi(m)$. Since the smoothing becomes larger for fainter magnitudes, the model automatically degrades the apparent slope of the TRGB edge as a function of distance.

The likelihood analysis works best if the red giant branch power--law
is observed over a range of one magnitude or more. In cases in which we
failed to detect the TRGB, we seldom observed a power law slope for the red giant branch,
with crowding and incompleteness effects leading to a rolloff of detected 
stars that did not fit the TRGB form.  
In these cases our model for $g(m)$ is inadequate and the likelihood
analysis is meaningless.  In any event, it is necessary to set the limits
of the magnitude range for which the likelihood analysis is performed;  it is
critical to ensure that the faint limit avoids regions where
incompleteness is non-negligible.  The bright limit is less critical, but it obviously must be set brighter than the TRGB. We find that setting it at $0.5$ or more magnitudes brighter than the TRGB gives good results.

\subsection{\bf Uncertainty in measuring the TRGB}

We have used a different approach for estimating random error in the TRGB magnitude from most previous authors. Errors from the method of SMF96 are generally obtained from the Sobel edge--detection response. The uncertainty of $I_{\rm TRGB}$ is taken as either the $\sigma$ in a Gaussian fit to the peak response or the FWHM of the peak, e.g. \citet{sakai00}. Instead we use bootstrap resampling of the data \citep[and references therein]{babu96} to simulate the act of making the same observation multiple times. This allows for the most realistic estimate of the standard deviation in the mean.

The magnitudes of all the stars brighter than a given limiting magnitide $M_l$ in an entire population have some luminosity function, $\Phi (M > M_l)$. We observe only a random fraction of the whole stellar population and so determine an observed luminosity function, $\hat{\Phi}(M > M_l)$. From this luminosity function we determine $I_{\rm TRGB}$ via either edge detection or maximum likelihood analysis; let us simply refer to the result of either process as ${\rm TRGB}(\hat{\Phi})$. We then randomly resample, with replacement, all of the observed magnitudes (i.e., any given star may appear more than once or not at all). Each stellar magnitude, $M_i$, carries with it some magnitude error, $\epsilon_i$. Thus, during the resampling $M_i$ is perturbed by its errors such that the resampled magnitude, $M_i^*$, will be randomly drawn from a Gaussian distribution about the original magnitude with $\epsilon_i$ as its standard deviation. For each resampling a bootstrap luminosity function , $\hat{\Phi}^*$, is then determined, and $I_{\rm TRGB}^*$ is measured via ${\rm TRGB}(\hat{\Phi}^*)$. We perform the bootstrap resampling 500 times for each galaxy and take the standard deviation of the distribution of $I_{\rm TRGB}^*$ as the  uncertainty in the measurement of $I_{\rm TRGB}$. We find that in most cases the estimated uncertainty is a few hundredths to one tenth of one magnitude. Details for each galaxy are given in \S~4.3.

\section{\bf Distances to Local Galaxies}

\subsection{\bf TRGB Calibration}

With a robust measurement of the $I$ magnitude of the TRGB, $I_{\rm TRGB}$, we may now calculate the distance modulus. We start with the following equation:
\begin{equation}
(m - M)_{I} = I_{\rm TRGB} + BC_{I} - M_{\rm bol, TRGB},
\end{equation}
where 
$M_{\rm bol, TRGB}$ is the bolometric magnitude of the TRGB and $BC_{I}$ is the bolometric correction to the $I$ magnitude of the TRGB.

The quantity $M_{\rm bol, TRGB}$ was found by \citet[hereafter referred to as DA90]{dacosta90} empirically to be a function of the metallicity [Fe/H]:
\begin{equation}
M_{\rm bol, TRGB} = -0.19[{\rm Fe/H}] - 3.81 .
\end{equation}
They also found that $BC_{I}$ is a function of $(V - I)$ color,
\begin{equation}
BC_{I} = 0.881 - 0.243(V - I)_{\rm TRGB}.
\end{equation}
DA90 determined an empirical relation between $(V - I)$ color and metallicity, but \citet{lee93} found it necessary to alter the relation slightly for more distant (and hence fainter) systems, obtaining
\begin{equation}
{\rm [Fe/H]} = -12.64 + 12.6(V - I)_{-3.5} - 3.3(V - I)_{-3.5}^{2} ,
\end{equation}
where $(V - I)_{-3.5}$ is the dereddened red giant branch $(V - I)$ color measured at an absolute $I$ magnitude of $-3.5$. The absolute magnitude of the TRGB is given by
\begin{equation}
M_{I, {\rm TRGB}} = M_{\rm bol, TRGB} - BC_{I}.
\end{equation}
To determine $M_{I, {\rm TRGB}}$ we iteratively calculate the distance modulus and the metallicity until they converge. The amount of Galactic extinction (and hence reddening) is calculated for each galaxy using the $I$ and $V$ extinctions given in the NASA/IPAC Extragalactic Database (NED), which are based upon the dust maps of \citet{schlegel98}.

To measure the $(V-I)$ color of the TRGB, we calculate a histogram of the colors of stars within $I_{\rm TRGB} \pm 0.1$ mag. We then fit a Gaussian to that histogram (binned at 0.01 mag) and take its peak as a first guess for $(V-I)_{\circ, {\rm TRGB}}$. To determine the uncertainty in this value we carry out a bootstrap resampling procedure similar to that used in determination of the uncertainty in $I_{\rm TRGB}$. Stellar magnitudes and colors are randomly resampled (with replacement) and perturbed by their internal random errors. The histogram calculation and Gaussian fitting are then carried out for each bootstrap resampling (we used 800 bootstrap samples). We fit a Gaussian to the final distribution of $(V-I)_{\circ, {\rm TRGB}}$ values and use its peak and standard deviation as the best estimate of $(V-I)_{\circ, {\rm TRGB}}$ and its uncertainty. The same procedure is used to determine $(V - I)_{-3.5}$. These values are given for each galaxy in Table~\ref{table:dist}.

The calibration relations were derived over the metallicity range of the Galactic globular clusters ($ -2.1 \leq {\rm [Fe/H]} \leq -0.7$) and are expected to work well over ages spanning 2 - 15 Gyr. The zero point was calibrated to the same level as that of the RR Lyrae distance scale (used to estimate the distances of the Galactic globular clusters). \citet{lee93} states generally that for this range $M_{I, {\rm TRGB}} \approx -4.0 \pm 0.1$ mag. More specifically, using the above calibration relations and metallicity range one finds $M_{I, {\rm TRGB}} = -4.01$ mag with an rms uncertainty of $\pm 0.03$ mag and a systematic error of $\pm 0.18$ mag.
\citet{ferrarese00b} treated the TRGB as a secondary distance indicator and calibrated a zero point from galaxies with Cepheid distances that yielded $M_{I, {\rm TRGB}} = -4.06 \pm 0.07 ({\rm random}) \pm 0.13 ({\rm systematic})$ mag. Also, \citet{bellazzini01} determined a new TRGB calibration based on photometry and a distance estimate from a detached eclipsing binary in the Galactic globular cluster $\omega$ Centauri. They found $M_{I, {\rm TRGB}} = -4.04 \pm 0.12$ mag. For this paper we use the calibration of \citet{lee93}, but in paper II we will use the calibration of \citet{ferrarese00b} for consistency with other recent TRGB measurements. 

\subsection{\bf Total Error Budget}

The true distance modulus for each galaxy is computed as
\begin{equation}
\mu_{\circ} = (m - M)_I - A_I = I_{\rm TRGB} - M_{I ,{\rm TRGB}} - A_I .
\end{equation}
We consider total errors in distance to be of two forms for these data. There are internal errors that are specific to each galaxy and external systematic errors that are the same for all galaxies.
The internal uncertainties in our distances are determined as the sum in quadrature of random and internal systematic errors. The random errors in the distance moduli are taken as the quadrature sum of the random error in measuring $I_{\rm TRGB}$ (estimated via bootstrap resampling) and $M_{I ,{\rm TRGB}}$ as such
\begin{equation}
(\sigma_{\mu_{\circ} ,{\rm random}})^2 = (\sigma_{\rm TRGB, bootstrap})^2 + (\sigma_{BC_I ,{\rm random}})^2 + (\sigma_{M_{bol},{\rm random}})^2 .
\end{equation}
Uncertainties in $M_{I ,{\rm TRGB}}$ arise from the errors in measuring the $(V-I)_{\circ}$ color at $I_{\rm TRGB}$ and at $M_I = -3.5$ as described in the previous section. These errors are propagated through the calibration equations (11-14) :
\begin{equation}
\sigma_{BC_I ,{\rm random}} = 0.243 \sigma_{(V-I)_{0,{\rm TRGB}},{\rm bootstrap}} ,
\end{equation}
\begin{equation}
\sigma_{M_{bol},{\rm random}} = 0.19 \sigma_{\rm [Fe/H]} ,
\end{equation}
and
\begin{equation}
\sigma_{\rm [Fe/H]} = |12.6 - 6.6 (V-I)_{-3.5}| \sigma_{(V-I)_{-3.5},{\rm bootstrap}} .
\end{equation}

Internal systematic errors include the uncertainties in the aperture corrections, the normalization to the standard photometric system ({\em i.e.} the photometric zero points), and the foreground extinction and reddening . Thus, we have
\begin{equation}
(\sigma_{\mu_{\circ} ,{\rm int,sys}})^2 = (\sigma_{I,{\rm sys}})^2 + (\sigma_{M_{I,{\rm TRGB}},{\rm int,sys}})^2 + (\sigma_{A_I})^2 ,
\end{equation}
\begin{equation}
\sigma_{M_{I,{\rm TRGB}},{\rm int,sys}} = (0.243 - 0.19 (12.6 - 6.6 (V-I)_{-3.5})) \sigma_{(V-I),{\rm sys}} ,
\end{equation}
and
\begin{equation}
(\sigma_{(V-I),{\rm sys}})^2 = (\sigma_{I,{\rm sys}})^2 + (\sigma_{V,{\rm sys}})^2 + (\sigma_{E(V-I)})^2 .
\end{equation}
The systematic magnitude errors ($\sigma_{I,{\rm sys}}$ and $\sigma_{V,{\rm sys}}$) are the quadrature sums of the aperture and zero point corrections in each bandpass. The uncertainties in extinction ($A_I$) and reddening ($E(V-I) = (A_V - A_I)$) are taken to be 10\% 
of their values (as per \citet{schlegel98}).

The values of $\sigma_{\rm TRGB, bootstrap}$ range from $\pm 0.02$ mag (Sextans~B and UGC~06456 using the Maximum Likelihood Method) to $\pm 0.12$ mag (Leo~I and UGC~03755 using the Edge Detection Method). The values of $\sigma_{M_{I,{\rm TRGB}}}$ range from $\pm 0.02$ mag to $\pm 0.04$ mag. For galaxies without observations in $V$ we use the full range of color/metallicity under which the calibrations were determined. This yields an rms uncertainty of $\pm 0.03$ mag. 
The total internal uncertainties in the true distance moduli are thus
\begin{equation}
(\sigma_{\mu_{\circ},{\rm internal}})^2 = (\sigma_{\mu_{\circ} ,{\rm random}})^2 + (\sigma_{\mu_{\circ} ,{\rm int,sys}})^2,
\end{equation}
ranging from $\pm 0.04$ mag (Sextans~B and UGC~06456 using the Maximum Likelihood Method) to $\pm 0.13$ mag (Leo I and UGC~03755 using the Edge Detection Method). 

The principle identified external systematic errors in the determination of the distance moduli arise from the uncertainty in the RR Lyrae distance scale for globular clusters ($\pm 0.15$ mag; see \citet{sakai99}), and the uncertainty in the TRGB calibration relations ($\pm 0.11$ mag):
\begin{equation}
(\sigma_{\mu_{\circ}, {\rm ext,sys}})^2 = (\sigma_{M_{I,{\rm TRGB}}, {\rm ext,sys}})^2 = (\sigma_{\rm RRLyrae})^2 + (\sigma_{BC_I ,{\rm sys}})^2 + (\sigma_{M_{bol},{\rm sys}})^2 .
\end{equation}
\citet{dacosta90} give the dispersion in equation (12) as $\pm 0.057$ mag and the dispersion in equation (11) as $\pm 0.09$ mag. Thus, the total systematic error common to all our TRGB distance measurements is $\pm 0.18$ mag.

\subsection{\bf TRGB Distances}
In order to gain confidence in our photometry and TRGB detection methodology we observed a few local galaxies with Keck that have been studied quite extensively by previous authors, including Leo I, NGC~3109, and Sextans B. We find that our derived distance moduli for these galaxies are in excellent agreement with those of previous authors. In our HST dataset we were able to successfully detect the TRGB for three galaxies and possibly a fourth. The definite detections were UGC~07577, UGC~06456, and NGC~1313, and the possible detection was UGC~03755. There were several galaxies in our combined Keck/{\em HST} dataset, such as NGC~2366 and NGC~2683, for which we were unable to successfully determine distances . Based upon distance estimates by previous authors using various methods we believe that these galaxies are sufficiently far away that their red giant branch stars were not resolved in our data. A summary of TRGB distances and uncertainties is presented in Table~\ref{table:dist}. Color-Magnitude diagrams (CMDs), luminosity functions, and other details for these galaxies are now presented.
\footnote{Photometry data for the galaxies presented here are archived in the NED/LEVEL5 database at the following URL: http://nedwww.ipac.caltech.edu/level5/March02/Mendez/.}

\subsubsection{\bf Leo I}

\begin{figure}
\begin{center}
\includegraphics[height=6in]{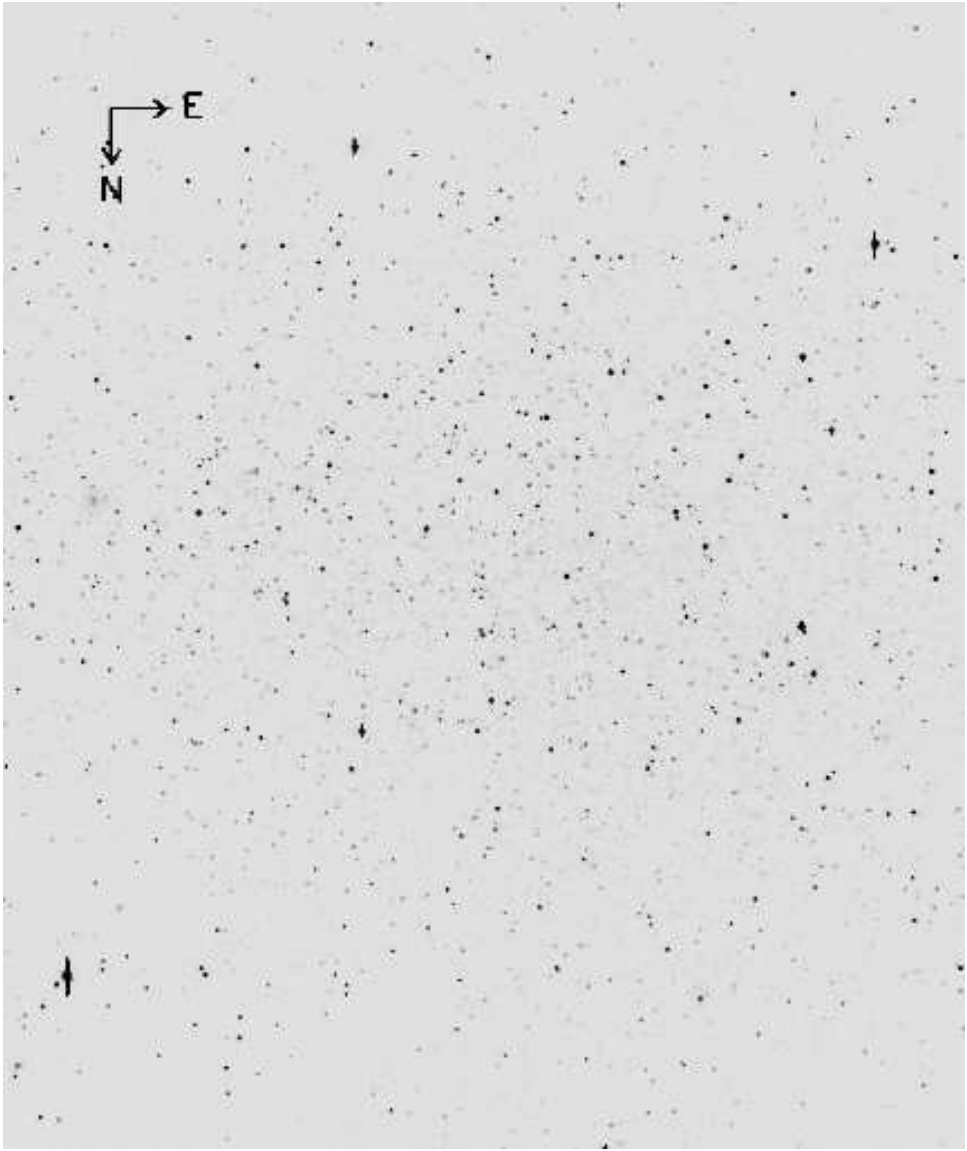}
\end{center}
\caption{Single 300s Keck/LRIS CCD image of Leo I in the $I$ band. The image is 5.5' x 7.3'.  }
\label{figure:leoi_image}
\end{figure}

\begin{figure}
\begin{center}
\includegraphics[height=8in]{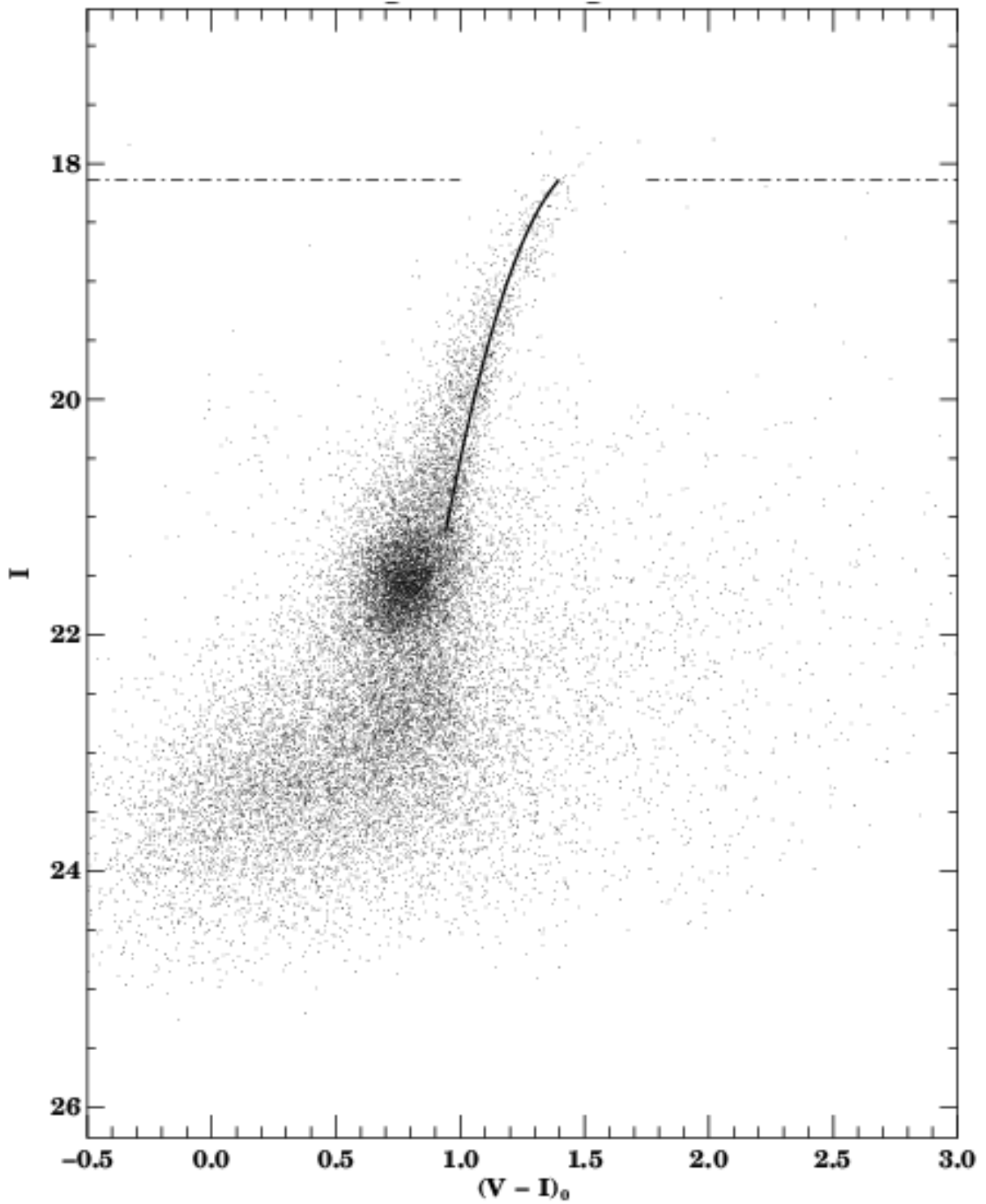}
\end{center}
\caption{Color--Magnitude Diagram for all stars in the Leo I CCD frame. The red giant branch locus (determined as discussed in the text) is also plotted. The $(V-I)$ color has been corrected for foreground reddening using the extinction values given in NED. }
\label{figure:leoi_cmd}
\end{figure}

The dwarf spheroidal galaxy Leo I is thought to be one of the most distant satellites of the Milky Way. Our observations with Keck were able to fully resolve Leo I into stars down to $I \sim 24$ mag (See Fig.~\ref{figure:leoi_image}). The Color-Magnitude diagram for the entire field reveals a very simple stellar population of red giants (Fig.~\ref{figure:leoi_cmd}). There appear to be almost no blue stars that may contaminate the luminosity function. The Red Giant Branch is extremely well defined with an unambiguous tip at $I = 18.14$ mag. Plotted in Fig.~\ref{figure:leoi_cmd} is an red giant branch locus for Leo I. We determine this locus by using the value of $(V-I)_{0 ,{\rm TRGB}}$ to interpolate between the two nearest of the four galactic globular cluster loci determined by DA90: M15, M2, NGC~1851, and 47 Tuc. The few stars above the tip are likely Asymptotic Giant Branch (AGB) stars. Also visible is the red clump of helium burning stars at $(V-I)_{\circ} \sim 0.5 - 0.9$ mag and $21.0 \leq I \leq 22.0 $ mag.

\begin{figure}
\begin{center}
\includegraphics[height=8in]{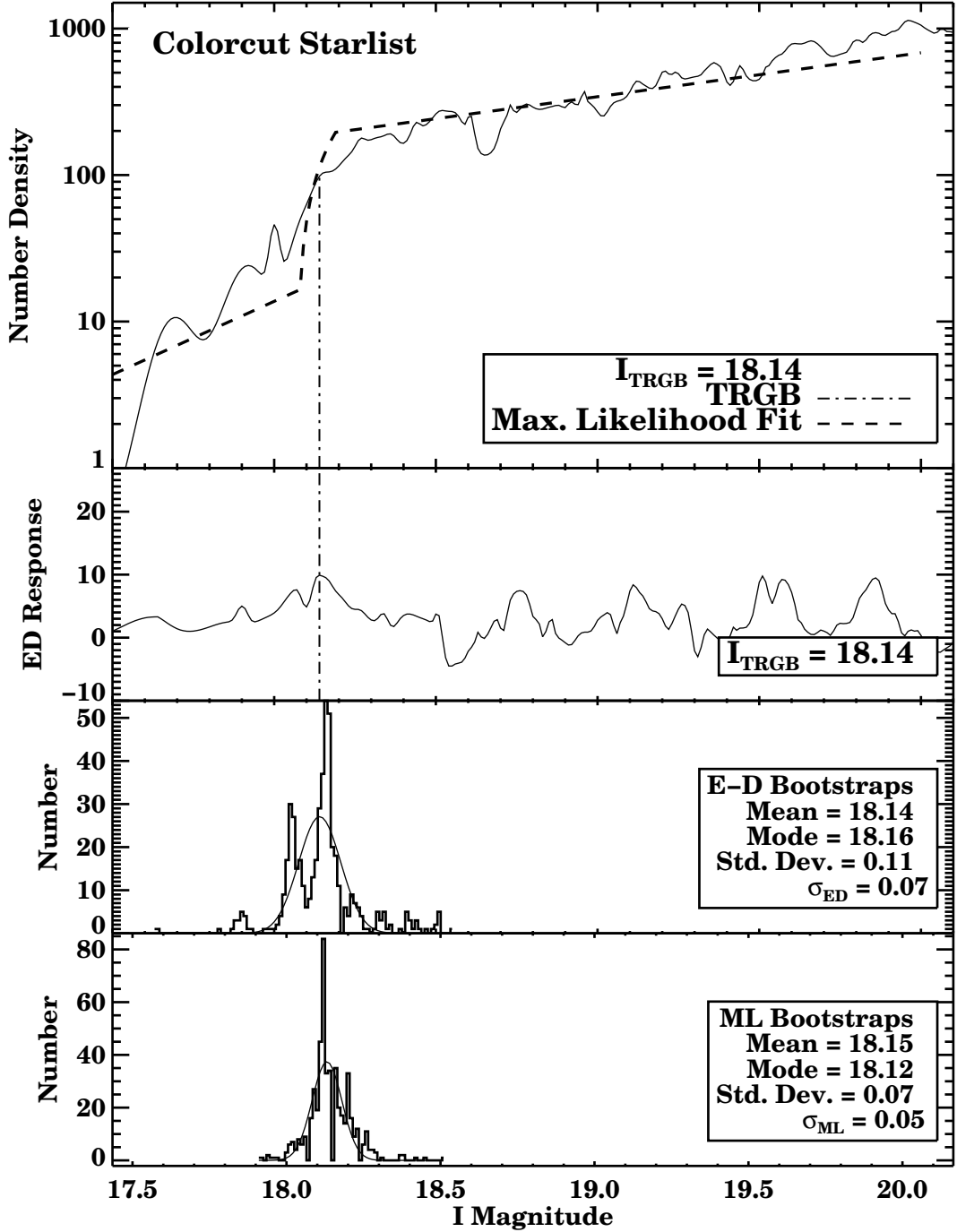}
\end{center}
\caption{Luminosity function of stars in Leo I with $0.3 \leq (V-I) \leq 1.5$ mag (top panel). Overplotted is the best-fit model luminosity function determined via maximum likelihood analysis. The second panel plots the weighted logarithmic edge--detection response; the dot--dash line indicates the TRGB location. The third and fourth panels plot the distribution of the bootstrap results for edge--detection and maximum likelihood methods, respectively.}
\label{figure:leoi_ccut_lf}
\end{figure}

\begin{figure}
\begin{center}
\includegraphics[height=8in]{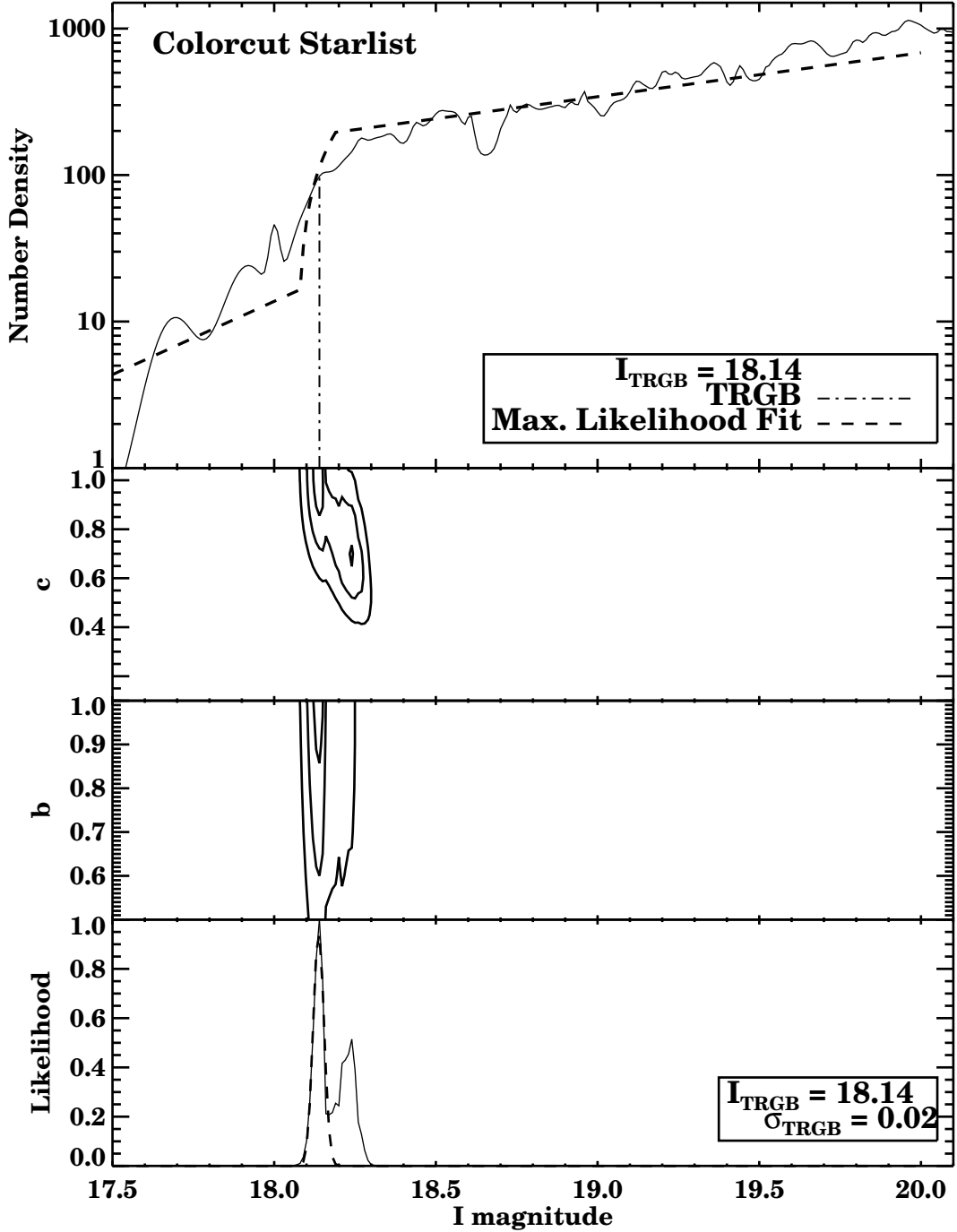}
\end{center}
\caption{Luminosity function of stars in Leo I with $0.3 \leq (V-I) \leq 1.5$ mag (top panel). Overplotted is the best-fit model luminosity function determined via maximum likelihood analysis. The second and third panels plot $1\sigma$, $2\sigma$, and $3\sigma$ contours of the likelihood function parameters $c$ and $b$, respectively. The fourth panel plots the marginalized likelihood. A Gaussian fit with $\sigma_{\rm TRGB} = 0.02$ mag is overplotted as a dashed line.}
\label{figure:leoi_ccut_like}
\end{figure}

Fig.~\ref{figure:leoi_ccut_lf} shows the logarithmic luminosity function for all stars with  $0.3 \leq (V-I) \leq 1.5$ mag. The upper red giant branch power law is quite obvious and has a clear break. Both the maximum likelihood model and the modified Sobel edge detector agree on the location of the break. For the maximum likelihood analysis we determine  $I_{\rm TRGB} = 18.14$ mag as shown in Fig~\ref{figure:leoi_ccut_like}. Bootstrap resampling gives an uncertainty of $\pm 0.07$ mag, with a distribution shown in Fig.~\ref{figure:leoi_ccut_lf}. The edge--detector also finds $I_{\rm TRGB} = 18.14$ mag with a bootstrap uncertainty of $\pm 0.11$ mag. These uncertainties are among the largest of the galaxies reported in this paper, despite Leo I being the closest by far. The large uncertainty is due to the small number of stars at the TRGB. With so few stars at the TRGB it is easier for it to move by a few hundredths of a magnitude with each bootstrap. \citet{madore95} studied the effect of small population sizes and estimated that if more than $\sim 100$ stars can be imaged in the first full magnitude interval then any systematic effect in the distance modulus would be well less than $0.1$ mag. There are just over 100 stars in the first magnitude of the luminosity function of Leo I.
Thus, using the method described in \S~4.1 we find $(V-I)_{\rm 0,TRGB} = 1.39 \pm 0.02$ mag and [Fe/H] = $-1.94 \pm 0.08$ dex. With a line of sight extinction of $A_I = 0.07$ mag we find $(m - M)_I - A_I = \mu_{\circ} = 22.05 \pm 0.10$(internal)$\pm 0.18$(systematic) mag for the maximum likelihood case. This is in excellent agreement with a previously published distance modulus for Leo I of $(m-M)_V = 22.00 \pm 0.15$ mag \citep{caputo99}. Another estimate of the TRGB for Leo~I was made by \citet{lee93b}, who found $I_{\rm TRGB} = 18.25 \pm 0.1$ mag (yielding $\mu_{\circ} = 22.18 \pm 0.11$ mag for the assumed $A_I = 0.04$ mag). The difference between their measurement of $I_{\rm TRGB}$ and that presented here is $0.11 \pm 0.13$ mag, again in good agreement.

\subsubsection{\bf Sextans B}

\begin{figure}
\begin{center}
\includegraphics[height=8in]{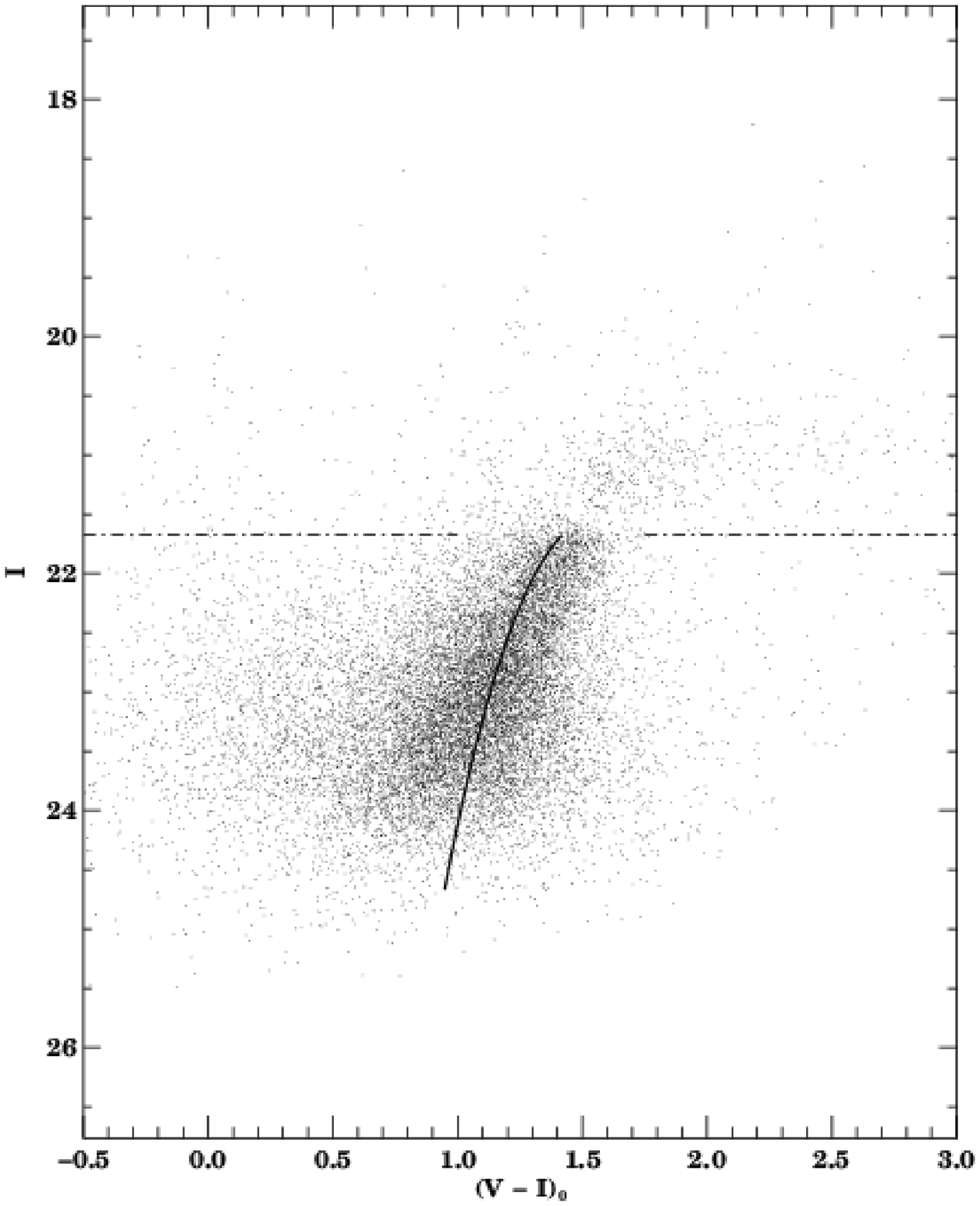}
\end{center}
\caption{Color--Magnitude Diagram for all stars in the Sextans B CCD frame. The red giant branch locus (determined as discussed in the text) is also plotted.}
\label{figure:SextansB_cmd}
\end{figure}

\begin{figure}
\begin{center}
\includegraphics[height=7.3in,angle=270]{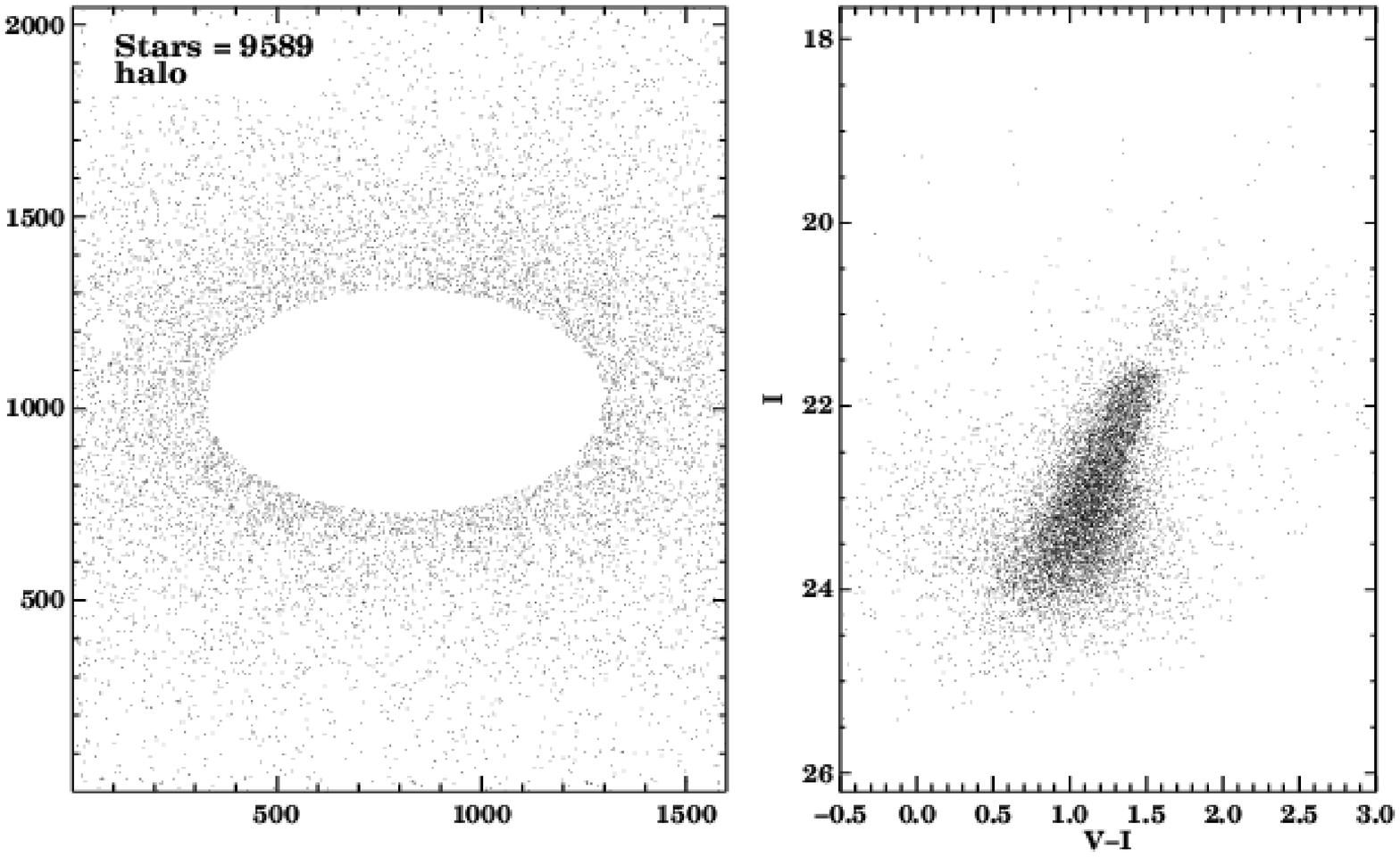}
\end{center}
\caption{The left panel is a map of the pixel positions of all stars in the halo region of Sextans B. The right panel is a Color--Magnitude diagram of just these stars. Notice that the TRGB is more obvious than in Fig.~\ref{figure:SextansB_cmd} after cutting out the core region.}
\label{figure:SextansB_halo_reg}
\end{figure}

Sextans B is a dwarf irregular galaxy located at the edge of the Local Group. It was well resolved in our Keck observations. Fig.~\ref{figure:SextansB_cmd} shows the CMD for the $\sim 20,000$ resolved stars with good photometry in the LRIS field. The stellar population appears fairly simple; the red giant branch is quite prominent and red giants clearly dominate the light of the galaxy. There is little contamination from blue stars. The stars above the red giant branch are likely AGB stars that do not obscure the conspicuous TRGB. In order to minimize these contaminants and those arising from crowding and photometric errors we examined the CMD of only the stars in an arbitrary halo region. Fig.~\ref{figure:SextansB_halo_reg} shows the positions of the 9500 stars in this halo region and their CMD. The TRGB was already obvious in Fig.~\ref{figure:SextansB_cmd} but it can now be seen more clearly.

\begin{figure}
\begin{center}
\includegraphics[height=8in]{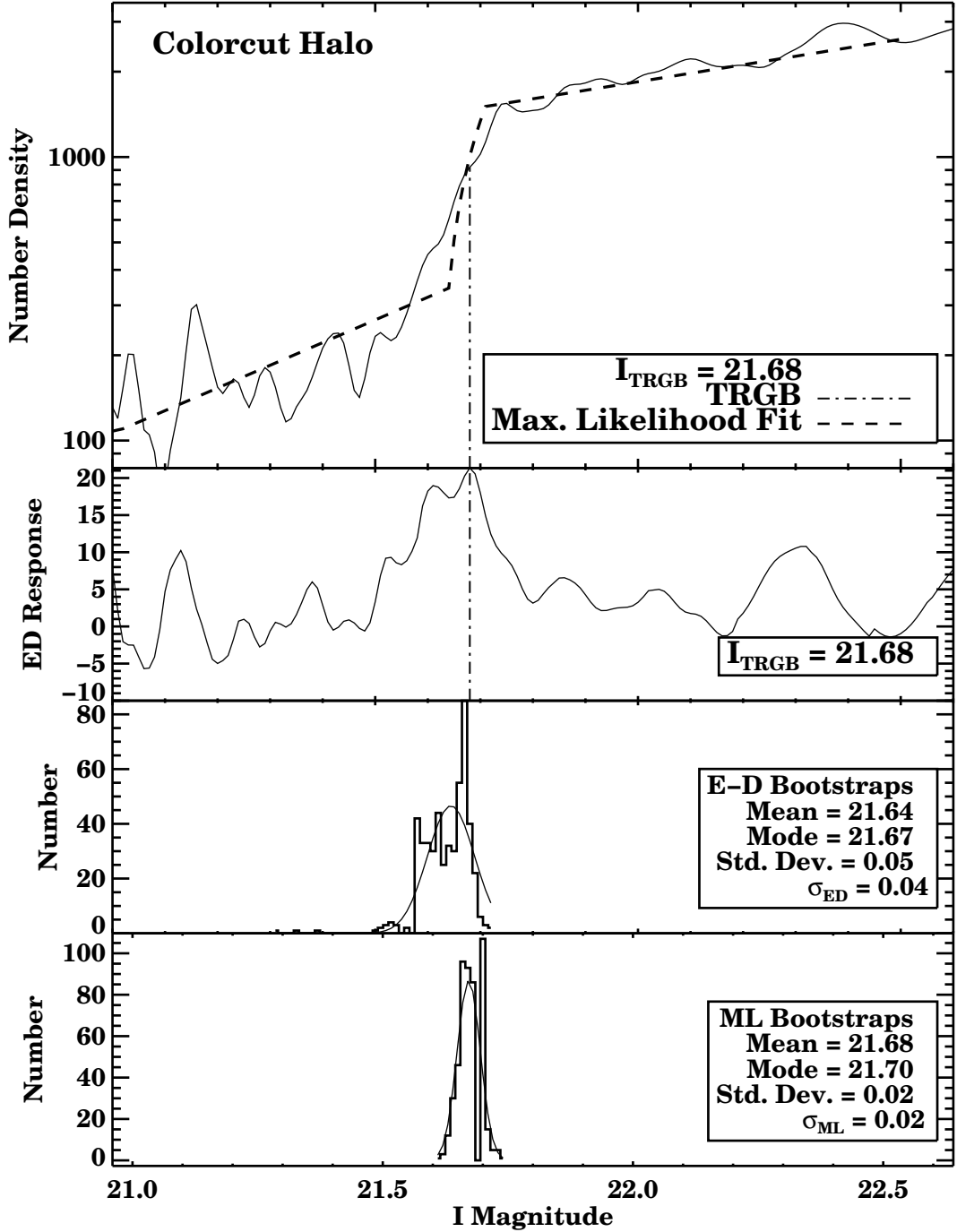}
\end{center}
\caption{Luminosity function of stars in the halo of Sextans B with $0.6 \leq (V-I) \leq 1.6$ mag (top panel). Overplotted is the best-fit model luminosity function determined via maximum likelihood analysis. The second panel plots the weighted logarithmic edge--detection response with a dot--dash line indicating the TRGB. The third and fourth panels plot the distribution of the bootstrap results for edge--detection and maximum likelihood methods, respectively.}
\label{figure:SextansB_halo_ccut_lf}
\end{figure}

Additionally, we make a color cut in the halo star list. Fig.~\ref{figure:SextansB_halo_ccut_lf} shows the resulting luminosity function of halo stars with $0.6 \leq (V-I) \leq 1.6$ mag. Again, the power-law distribution with index $a = 0.3$ is evident and has an unambiguous break at $I_{\rm TRGB} = 21.68$ mag in the maximum likelihood analysis as shown in Fig.~\ref{figure:SextansB_halo_ccut_like} (also $I_{\rm TRGB} = 21.68$ mag in the edge-detector). Bootstrap resampling of stars in the color-cut halo region gives an uncertainty of $\pm 0.02$ mag, shown in panels 3 and 4 of Fig.~\ref{figure:SextansB_halo_ccut_lf}. The TRGB color is estimated to be $(V-I)_{0 ,{\rm TRGB}} = 1.44 \pm 0.01$ mag and the metallicity is [Fe/H] = $-1.73 \pm 0.04$ dex. The \citet{schlegel98} dust maps give the foreground extinction as $A_I = 0.062$ mag. We assume that internal extinction is negligible as we are focusing on the halo region that is presumably mostly dust-free. Thus for the maximum likelihood method we determine a Population II distance modulus of $(m-M)_{\circ} = 25.63 \pm 0.04$(internal)$\pm 0.18$(systematic) mag corresponding to a heliocentric metric distance of $1.34 \pm 0.02$ Mpc. This result is in superb agreement with a previously published study of Sextans B by \citet{sakai97} using both the TRGB ($ (m - M)_{\circ} = 25.56 \pm 0.10$) and Cepheids ($(m - M)_{\circ} = 25.69 \pm 0.27$ mag).

\subsubsection{\bf NGC~3109}

\begin{figure}
\begin{center}
\includegraphics[height=8in]{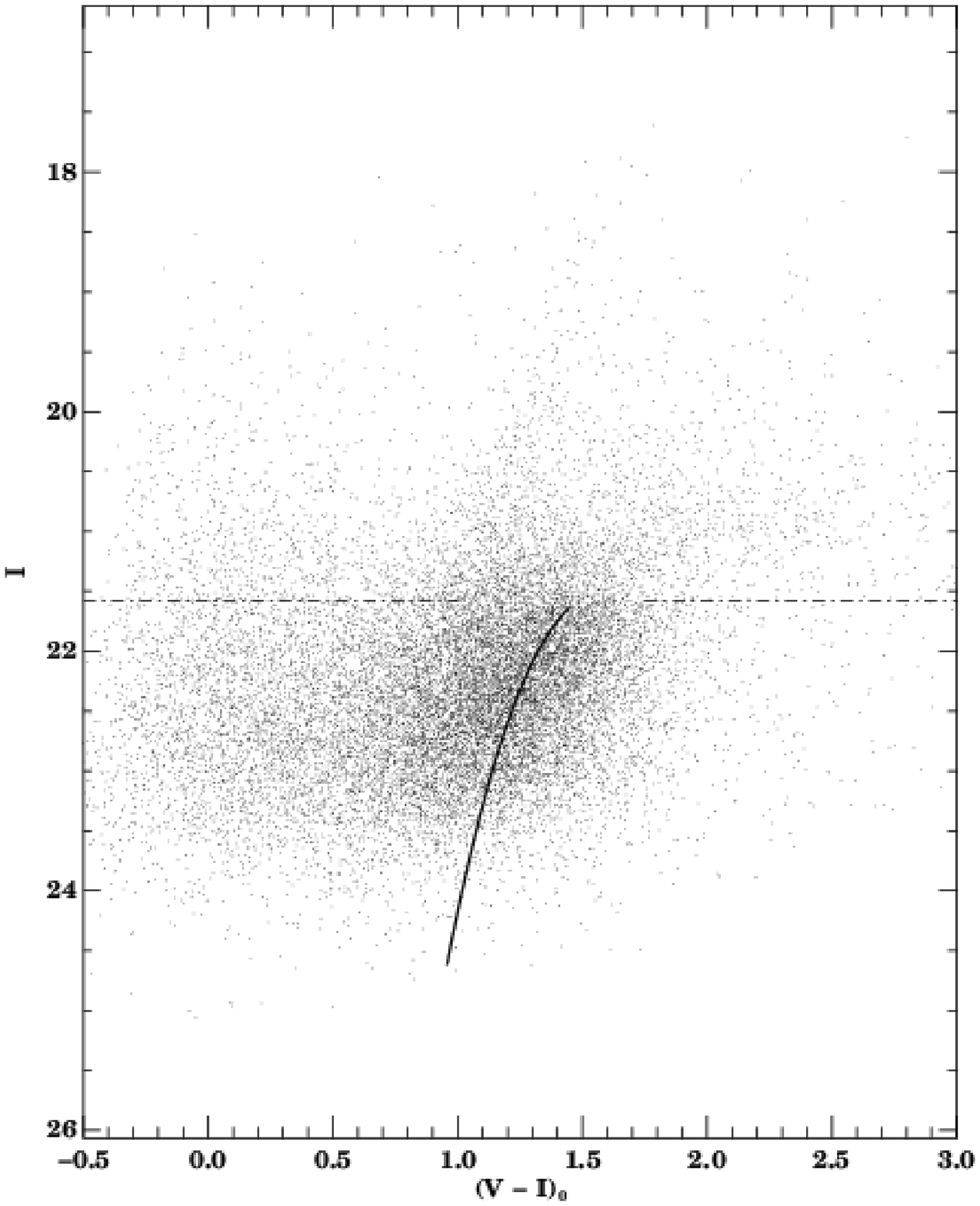}
\end{center}
\caption{Color--Magnitude Diagram for all stars in the NGC~3109 CCD frame. The red giant branch locus (determined as discussed in the text) is also plotted.}
\label{figure:NGC3109_cmd}
\end{figure}

NGC~3109 is another dwarf irregular galaxy, classified Sm IV, at the edge of the Local Group. It has a flattened disk region apparently viewed nearly edge-on. Our Keck observations have brilliantly resolved the galaxy into stars. In Fig.~\ref{figure:NGC3109_cmd} we present the CMD for the entire LRIS field. This CMD is considerably more complex than those of Sextans B or Leo I. A red giant branch is evident but there is large contamination from blue main sequence and asymptotic giant branch stars. There are also likely some red supergiants in the mix to further complicate matters. 

\begin{figure}
\begin{center}
\includegraphics[height=7.3in,angle=270]{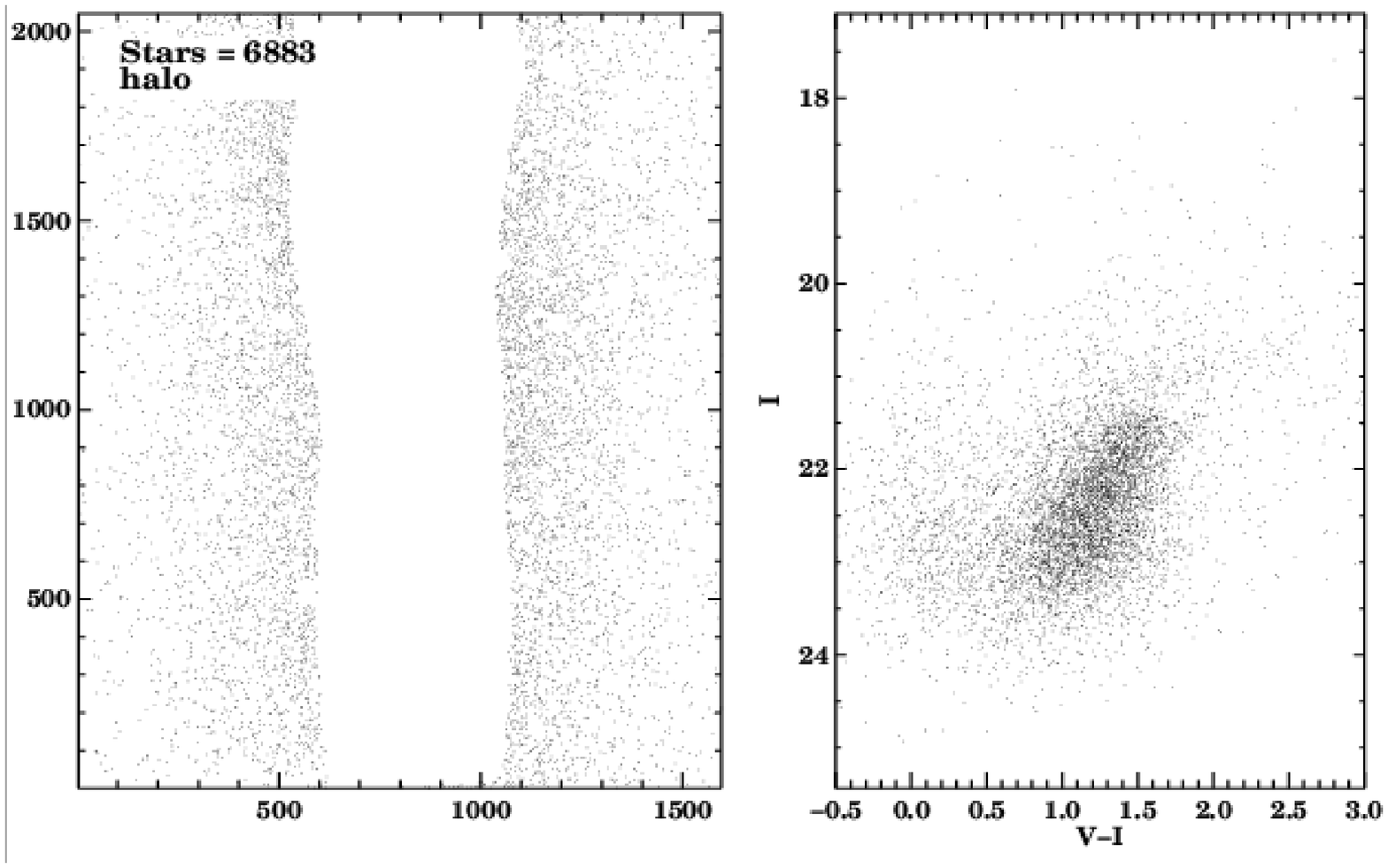}
\end{center}
\caption{The left panel is a map of the pixel positions of all stars in the halo region of NGC~3109. The right panel is a Color--Magnitude diagram of just these stars. Notice that the TRGB is more obvious than in Fig.~\ref{figure:NGC3109_cmd} after cutting out the core region.}
\label{figure:NGC3109_halo_reg}
\end{figure}

\begin{figure}
\begin{center}
\includegraphics[height=8in]{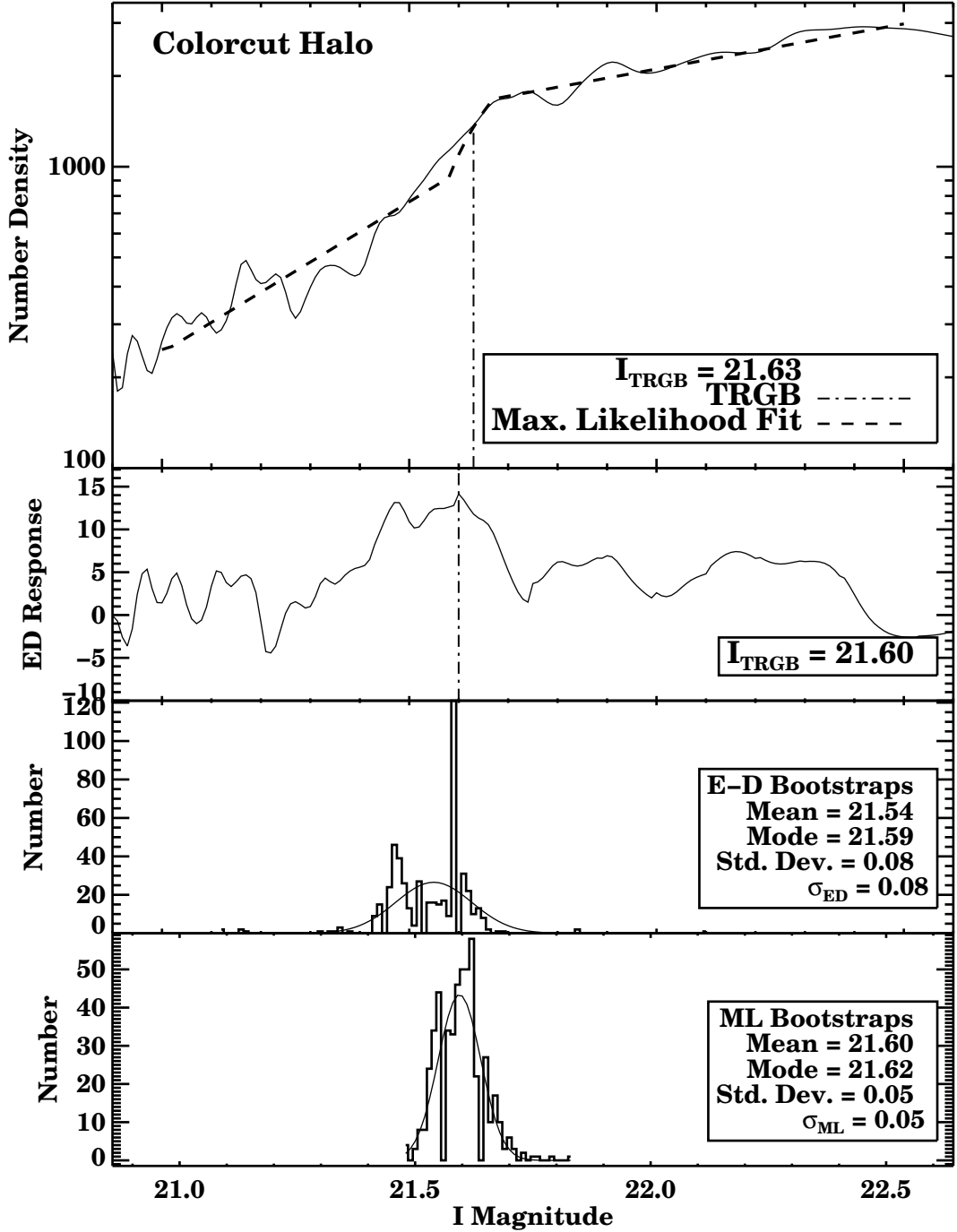}
\end{center}
\caption{Luminosity function of stars in the halo of NGC~3109 with $0.7 \leq (V-I) \leq 1.6$ mag (top panel). Overplotted is the best--fit model luminosity function determined via maximum likelihood analysis. The lower three panels are as in Fig.~\ref{figure:leoi_ccut_lf}}
\label{figure:NGC3109_halo_ccut_lf}
\end{figure}

\begin{figure}
\begin{center}
\includegraphics[height=8in]{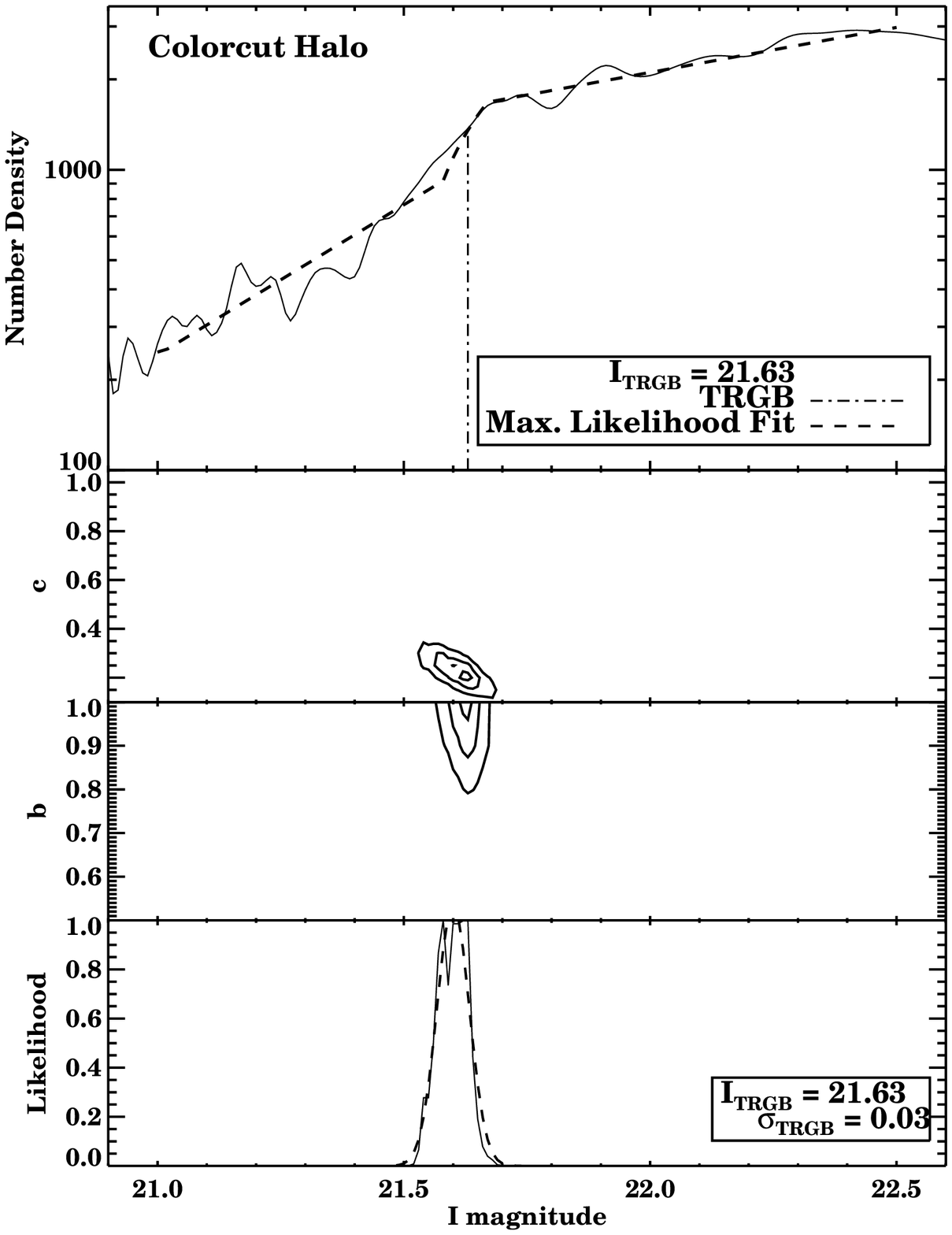}
\end{center}
\caption{Top panel is as in Fig.~\ref{figure:NGC3109_halo_ccut_lf}.  The second and third panels plot $1\sigma$, $2\sigma$, and $3\sigma$ contours of the likelihood function parameters $c$ and $b$. The fourth panel plots the marginalized likelihood. A Gaussian fit with $\sigma_{\rm TRGB} = 0.03$ mag is overplotted as a dashed line.}
\label{figure:NGC3109_halo_ccut_like}
\end{figure}

As we did with Sextans B, we selected stars from a halo region to reduce contamination around the TRGB. The pixel map and CMD for these stars is shown in Fig.~\ref{figure:NGC3109_halo_reg}. The red giant branch is now much more apparent as is its tip. The luminosity function of halo stars with $0.7 \leq (V-I) \leq 1.6$ mag is shown in Fig.~\ref{figure:NGC3109_halo_ccut_lf}. The two TRGB detection methods agree quite well and we find $I_{\rm TRGB} = 21.63 \pm 0.05$ mag for the maximum likelihood analysis, as shown in Fig.~\ref{figure:NGC3109_halo_ccut_like}, and $I_{\rm TRGB} = 21.60 \pm 0.08$ mag for the logarithmic edge--detector. The best estimate of the TRGB color is $(V-I)_{0 ,{\rm TRGB}} = 1.46 \pm 0.02$ mag, with [Fe/H] = $-1.69 \pm 0.06$ dex. For the maximum likelihood method and using an extinction value of $A_I = 0.129$ mag we find the distance modulus $(m-M)_{\circ} = 25.52 \pm 0.06$(internal)$\pm 0.18$(systematic) mag, which corresponds to a distance from the Sun of $1.27 \pm 0.04$ Mpc. This is within the errors of a previously published work on NGC~3109 \citep{minniti99} which found $(m - M)_{\circ} = 25.62 \pm 0.1$ mag using the tip of the red giant branch.

\subsubsection{\bf UGC~07577}

The dwarf irregular galaxy UGC~07577 was one of our {\em HST} target galaxies. Based on our experience with the Keck data we decided to minimize {\em HST} observation time by only obtaining images in the FW814 filter. This was done for galaxies not expected to have considerable contamination from blue main sequence or asymptotic giant branch populations. Even a large AGB population is not likely to affect the TRGB detection severely as it would surely have a different power-law slope in $I$ band from the red giant branch, so the break in the red giant branch power-law at the TRGB would remain.

\begin{figure}
\begin{center}
\includegraphics{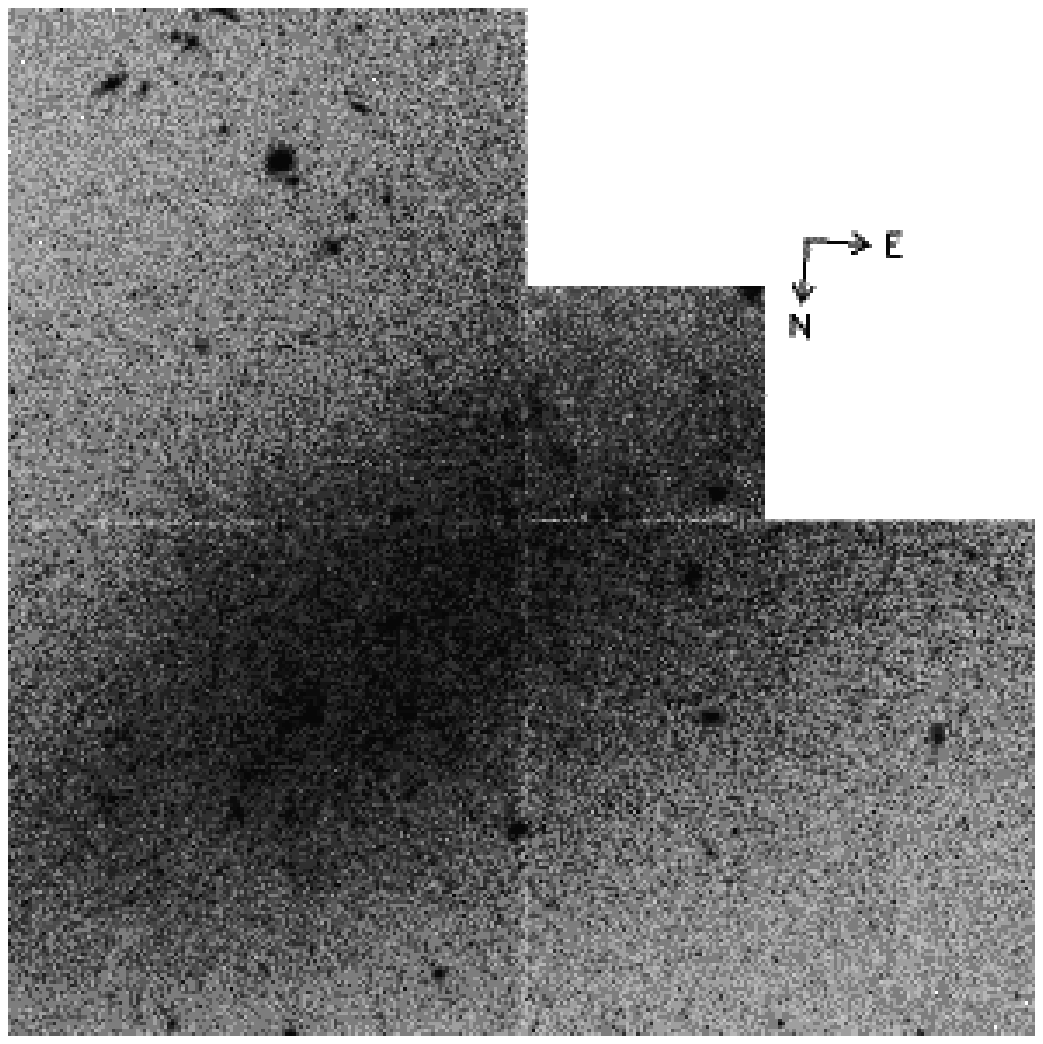}
\end{center}
\caption{Mosaic of a single 1300s {\em HST}/WFPC2 exposure of UGC~07577. }
\label{figure:ugc07577_image}
\end{figure}

\begin{figure}
\begin{center}
\includegraphics[height=8in]{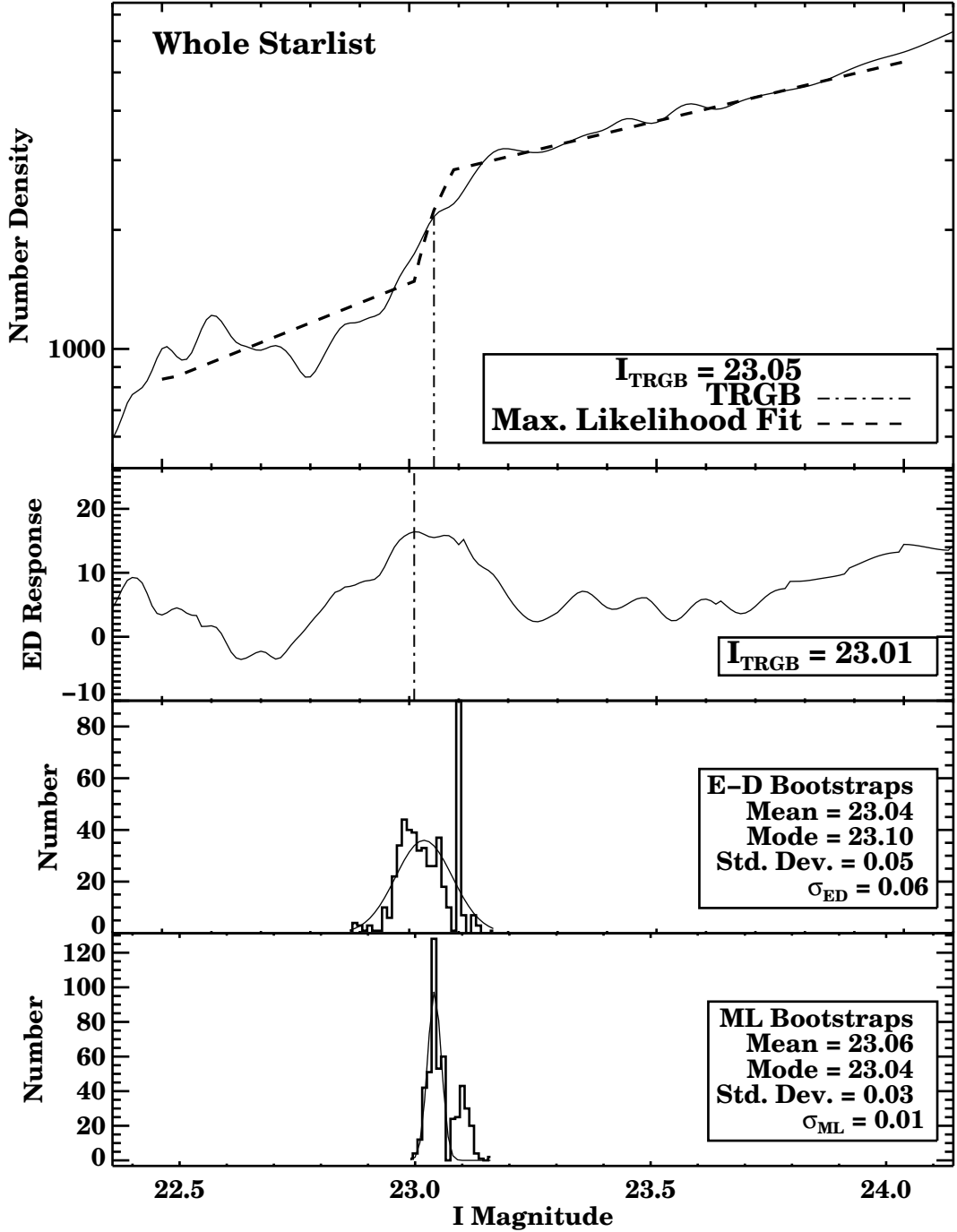}
\end{center}
\caption{Luminosity function of all stars in the {\em HST}/WFPC2 image of UGC~07577 (top panel). Overplotted is the best-fit model luminosity function determined via maximum likelihood analysis. The lower three panels are as in Fig.~\ref{figure:leoi_ccut_lf}}
\label{figure:ugc07577_lf}
\end{figure}

\begin{figure}
\begin{center}
\includegraphics[height=8in]{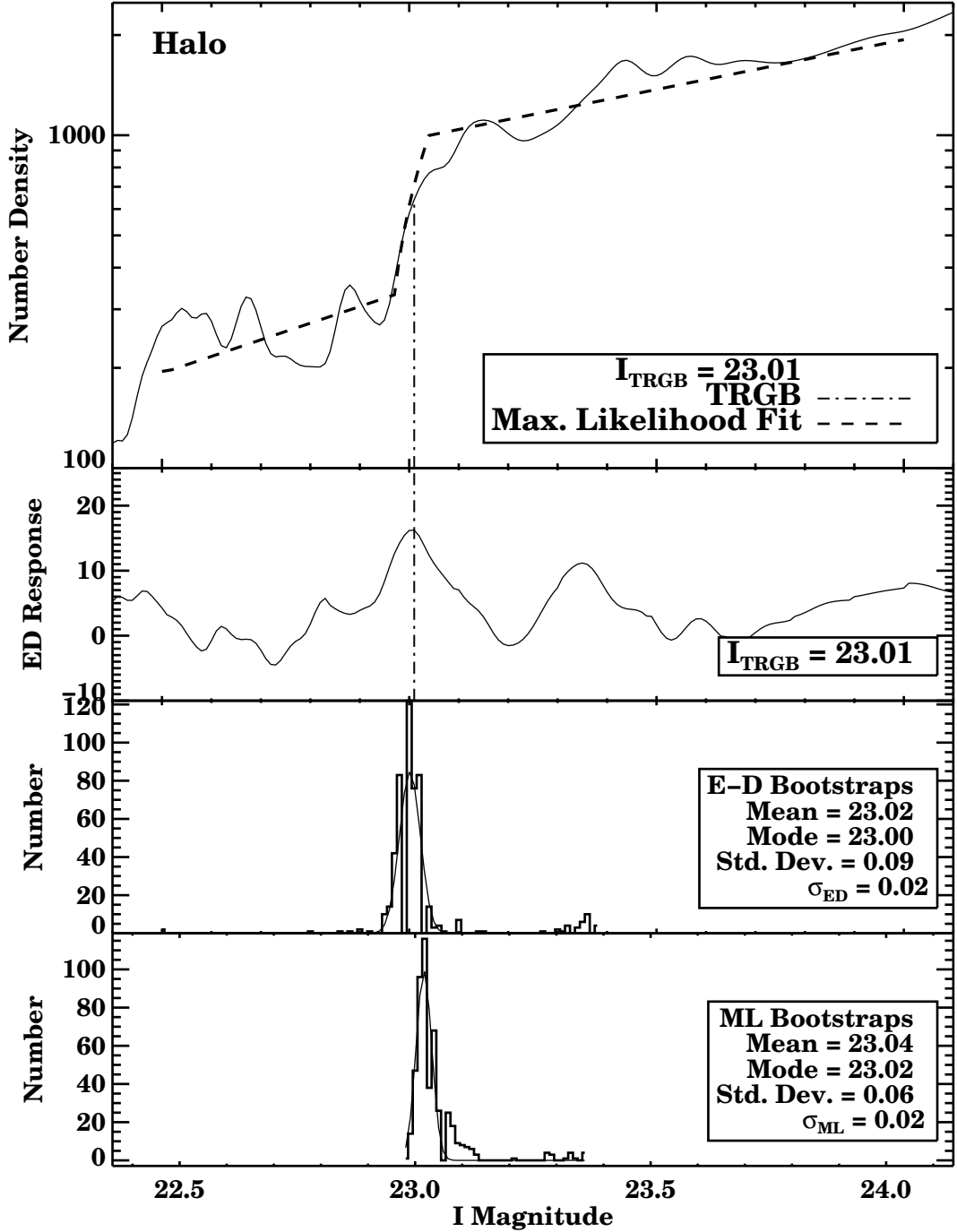}
\end{center}
\caption{Luminosity function of stars in the halo region of UGC~07577 (top panel). Overplotted is the best-fit model luminosity function determined via maximum likelihood analysis. The lower three panels are as in Fig.~\ref{figure:leoi_ccut_lf} }
\label{figure:ugc07577_halo_lf}
\end{figure}

\begin{figure}
\begin{center}
\includegraphics[height=8in]{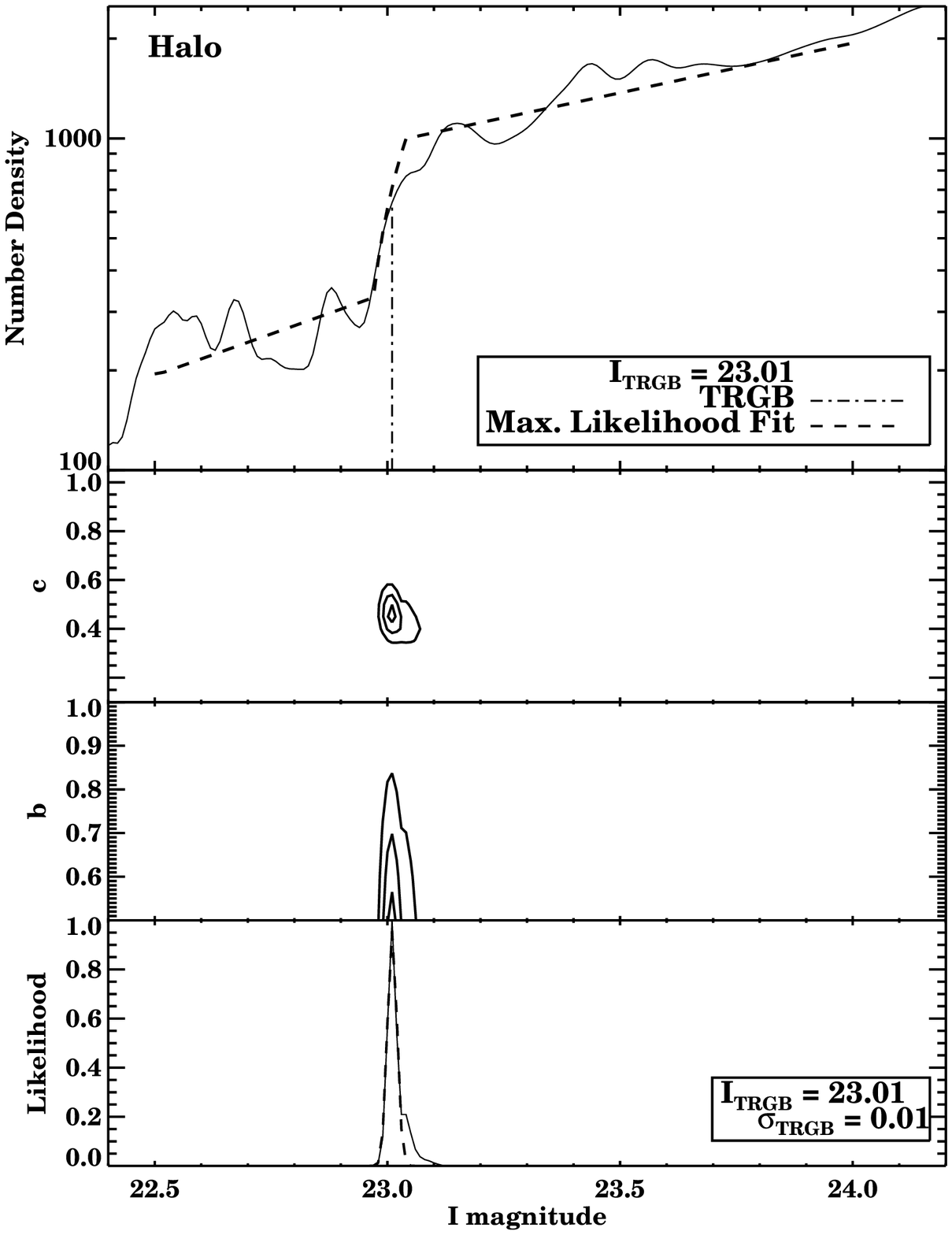}
\end{center}
\caption{Top panel is as in Fig.~\ref{figure:ugc07577_lf}. The second and third panels plot $1\sigma$, $2\sigma$, and $3\sigma$ contours of the likelihood function parameters $c$ and $b$. The fourth panel plots the marginalized likelihood. A Gaussian fit with $\sigma_{\rm TRGB} = 0.01$ mag is overplotted as a dashed line.}
\label{figure:ugc07577_halo_like}
\end{figure}

The galaxy was well resolved into stars (see Fig.~\ref{figure:ugc07577_image}). The galaxy (transformed into standard Cousins $I$) was broken up into core and halo regions as was done for the Keck objects. The luminosity functions of the entire field and of the halo are shown in Fig.~\ref{figure:ugc07577_lf}, and Fig.~\ref{figure:ugc07577_halo_lf} respectively. All show the same familiar red giant branch power-law slope, $a = 0.3$. The likelihood function for the entire field is maximized at $I_{\rm TRGB} = 23.05 \pm 0.03$ mag and the edge detector gives $I_{\rm TRGB} = 23.01 \pm 0.09$ mag. For the halo region the likelihood function is maximized at $I_{\rm TRGB} = 23.01 \pm 0.06$ mag as shown in Fig.~\ref{figure:ugc07577_halo_like} and this is the value we adopt. We assume that the mean metallicity of the UGC~07577 red giants is within the calibrated range of the Galactic globular clusters. Thus we adopt an absolute magnitude of $M_{\rm I,TRGB} = -4.01 \pm 0.03$ mag. Using the maximum likelihood value and an extinction of $A_I = 0.039$ mag we derive a distance modulus for UGC~07577 of $(m-M)_{\circ} = 26.98 \pm 0.05$(internal)$\pm 0.07$(systematic) mag, which is a heliocentric distance of $2.49 \pm 0.08$ Mpc. UGC~07577 has been little studied previously, but one recent estimate by \citet{tikhonov98} gives the distance as 4.8 Mpc. This is extremely discrepant, However they were estimating the distance using the magnitudes of the brightest blue super giant stars. Several authors have commented on the errors associated with this method, e.g. \citet{rozanski94,karachentsev94,lyo97}. While there have been differences between their estimates the errors are generally uncomfortably large ($\sim 0.79$ mag for the brightest blue stars).

\subsubsection{\bf NGC~1313}

\begin{figure}
\begin{center}
\includegraphics{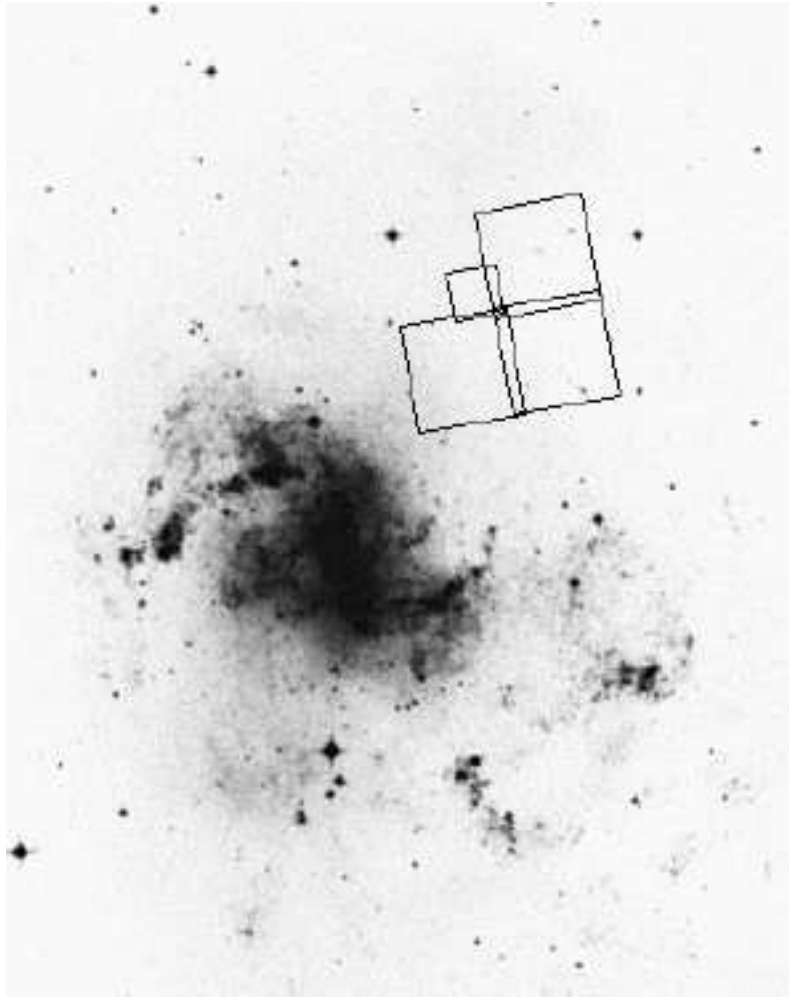}
\end{center}
\caption{{\em HST}/WFPC2 footprint of our observations overlayed on a Digital Sky Survey image of NGC~1313.}
\label{figure:ngc1313_dss_overlay}
\end{figure}

NGC~1313 is classified as a SB(s)d barred spiral and was described by \citet{devaucouleurs63} as a transition object from the normal barred spirals to the irregular Magellanic type barred spirals. It has considerable ongoing star formation and an active nucleus. Certainly the mixed stellar populations and the internal extinction of the disk are cause for concern in making a clear distance measurement based on the TRGB. For these reasons we obtained {\em HST} observations in the halo region of this galaxy 3.67 arcmin from its center (see Fig.~\ref{figure:ngc1313_dss_overlay}). In this region we expect the dominant stars to be Population II red giants. 

\begin{figure}
\begin{center}
\includegraphics[height=8in]{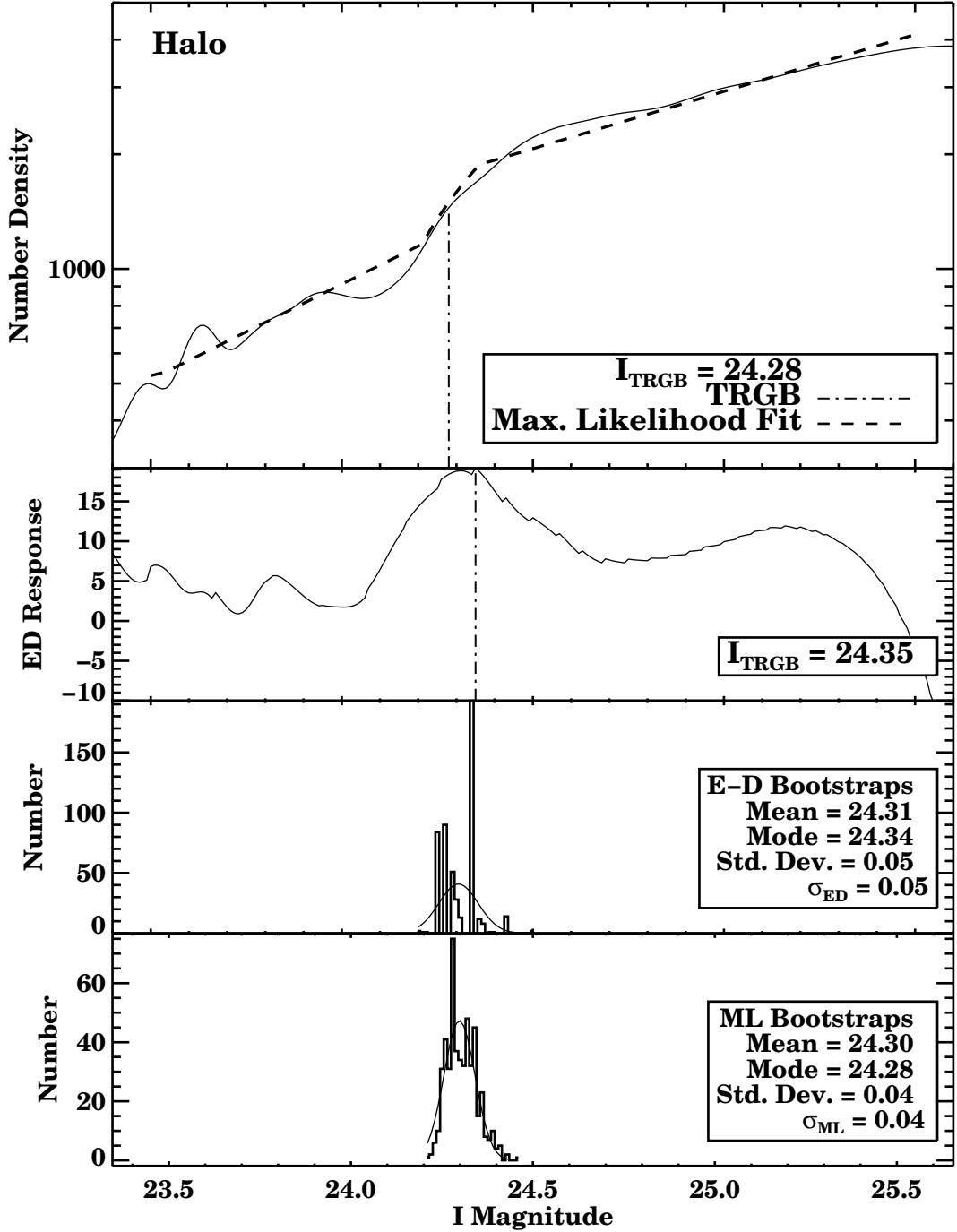}
\end{center}
\caption{Luminosity function of stars in the halo region of NGC~1313. Overplotted is the best-fit model luminosity function determined via maximum likelihood analysis. The lower three panels are as in Fig.~\ref{figure:leoi_ccut_lf} }
\label{figure:ngc1313_halo_lf}
\end{figure}

\begin{figure}
\begin{center}
\includegraphics[height=8in]{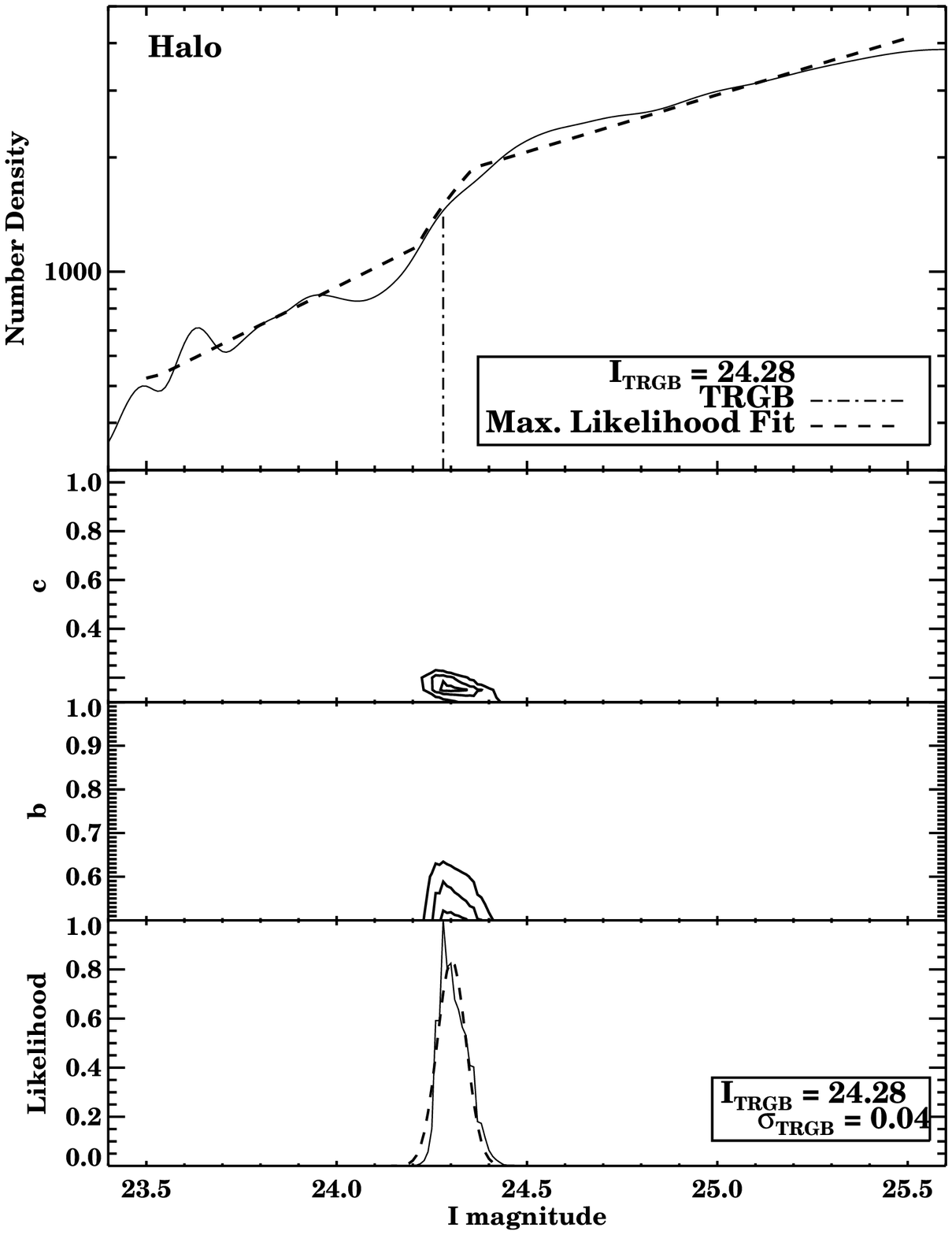}
\end{center}
\caption{Top panel is as in Fig.~\ref{figure:ngc1313_halo_lf}. The second and third panels plot $1\sigma$, $2\sigma$, and $3\sigma$ contours of the likelihood function parameters $c$ and $b$. The fourth panel plots the marginalized likelihood. A Gaussian fit with $\sigma_{\rm TRGB} = 0.04$ mag is overplotted as a dashed line.}
\label{figure:ngc1313_halo_like}
\end{figure}

We obtained images only in the F814W bandpass for NGC~1313 as we expected little contamination from other stellar types in this region of the galaxy. In Fig.~\ref{figure:ngc1313_halo_lf} we present the $I$ band luminosity function for the entire WFPC2 field of view. The tell-tale power-law signature ($a = 0.3$) of the red giant branch is visible from $I \sim 24.3$ mag down to the photometric limits at $I = 25.5$ mag. While there are many stars brighter than $I \sim 24.3$ mag they would appear to cause little doubt that the red giant branch terminates there. The maximum likelihood analysis gives the magnitude of the TRGB for NGC~1313 as $I_{\rm TRGB} = 24.28 \pm 0.04$ mag, as shown in Fig.~\ref{figure:ngc1313_halo_like}, and the modified Sobel edge--detector gives $I_{\rm TRGB} = 24.35 \pm 0.05$ mag. Since we do not have color information for the NGC~1313 halo we assume that its metallicity is within the calibrated range and thus use $M_{\rm I,TRGB} = -4.01 \pm 0.03$ mag. Using the maximum likelihood value and the \citet{schlegel98} extinction of $A_I = 0.212$ mag we obtain a Population II distance modulus for NGC~1313 of $(m-M)_{\circ} = 28.08 \pm 0.06$(internal)$\pm 0.18$(systematic) mag corresponding to a heliocentric distance of $4.13 \pm 0.11$ Mpc.  Using the magnitudes of brightest resolved stars, \citet{devaucouleurs63} estimated the distance of NGC~1313 at 4.5 Mpc. \citet{tully88} additionally gives the distance to NGC~1313 as 3.7 Mpc. 

\subsubsection{\bf UGC~06456}

In addition to our own {\em HST} data we searched through the {\em HST} archive at STScI for WFPC2 observations of galaxies in our all sky sample in at least F814W. While we found several good candidates, most of the observations we found were for disk regions within spiral galaxies. We have found at least one excellent dataset to add to our study. The observations (program 6276, PI J. Westphal) were taken as part of a project to study peculiar and interacting galaxies. From the archive data we used three 1400s exposures in F555W and three 1400s exposures in F814W of the blue compact dwarf galaxy (BCD) UGC~06456. We reduced these data as described in \S~2.2 for our own {\em HST} observations.

\begin{figure}
\begin{center}
\includegraphics[height=8in]{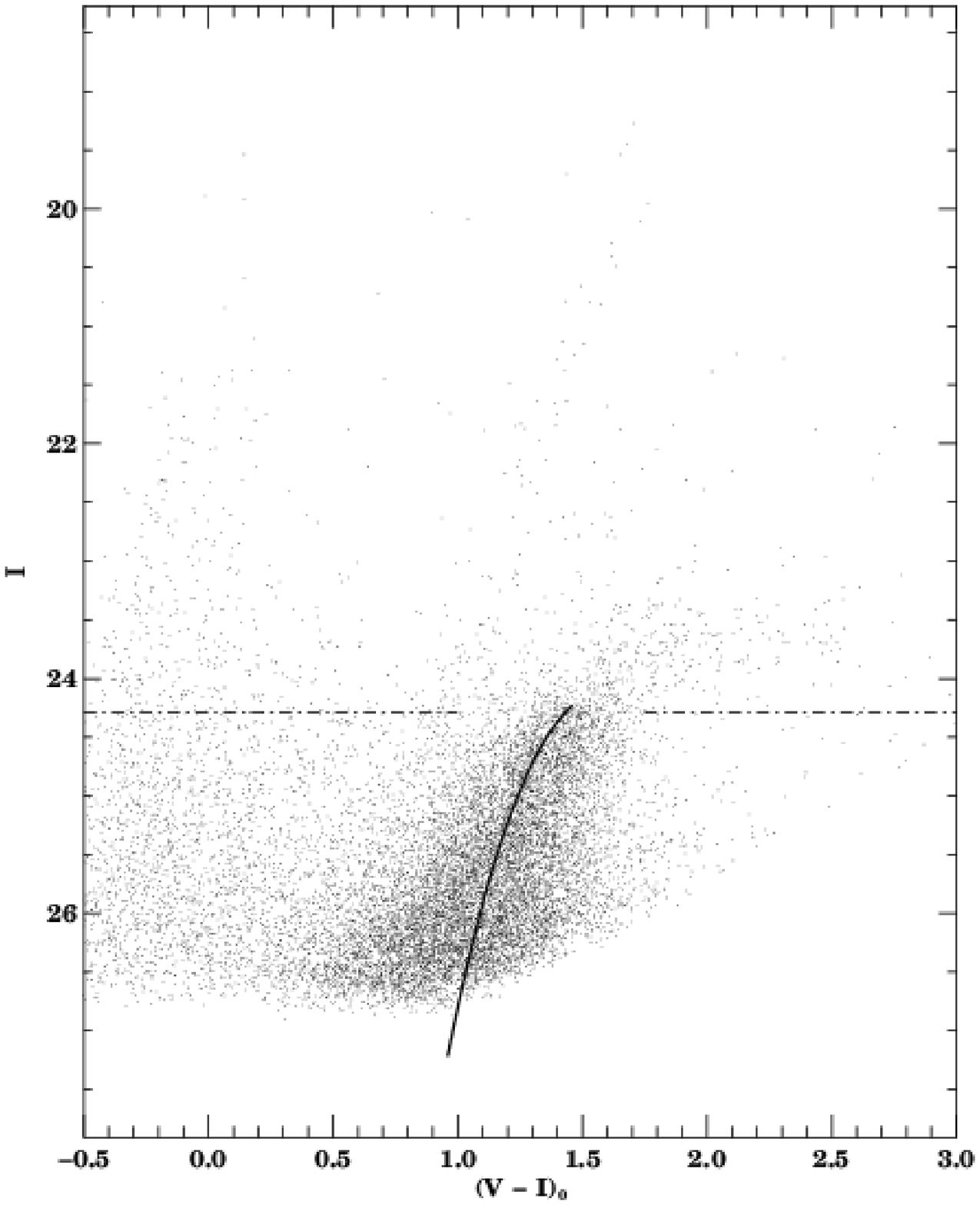}
\end{center}
\caption{Color--Magnitude Diagram for all stars in the UGC~06456 CCD frame. The red giant branch locus (determined as discussed in the text) is also plotted.}
\label{figure:ugc06456_cmd}
\end{figure}

\begin{figure}
\begin{center}
\includegraphics[height=7.3in,angle=270]{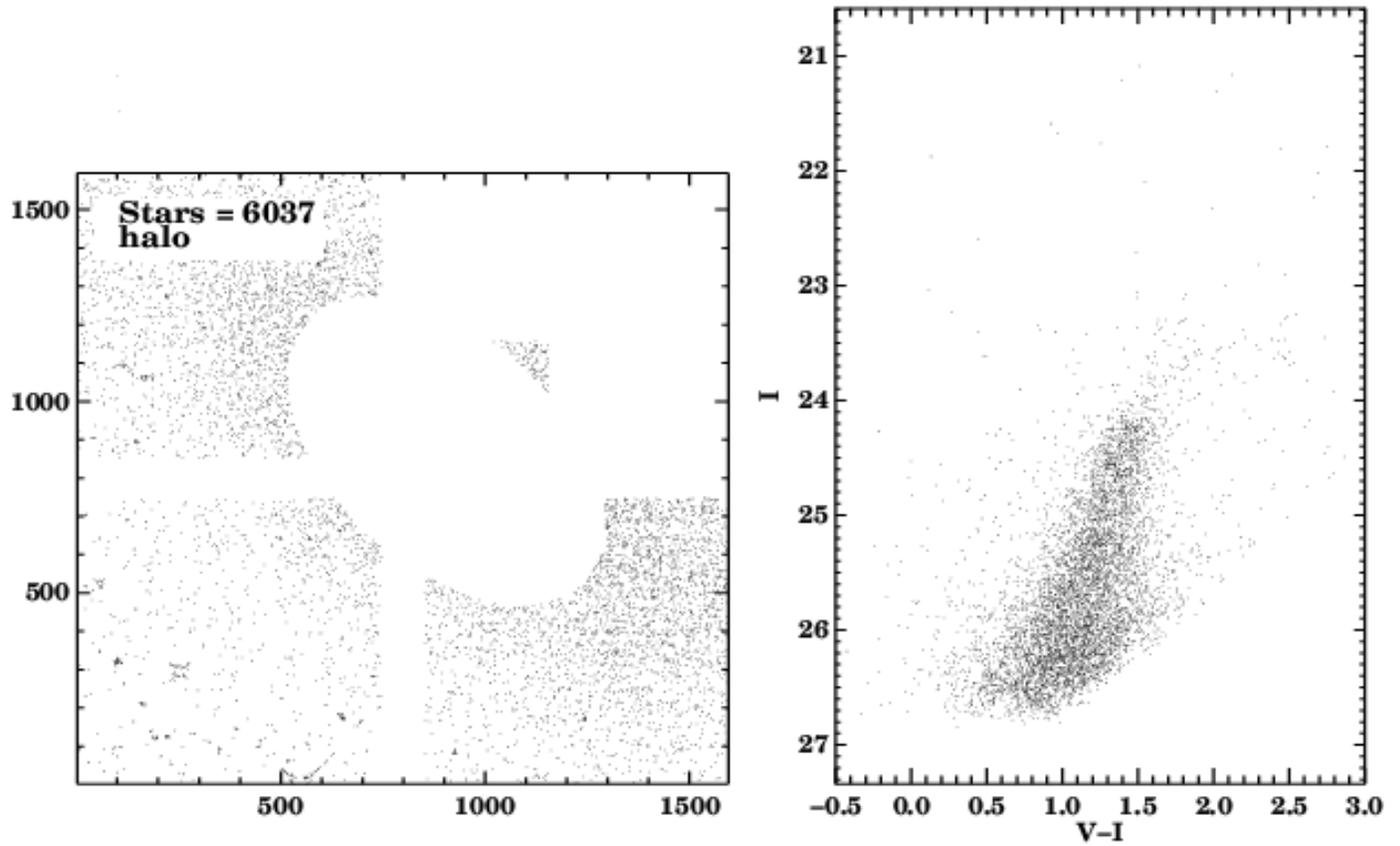}
\end{center}
\caption{The left panel is a map of the pixel positions of all stars in the halo region of UGC~06456. The right panel is a Color--Magnitude diagram of just these stars. Notice the considerable drop in the number of blue stars from Fig.~\ref{figure:ugc06456_cmd} after cutting out the core region.}
\label{figure:ugc06456_halo_reg}
\end{figure}

\begin{figure}
\begin{center}
\includegraphics[height=8in]{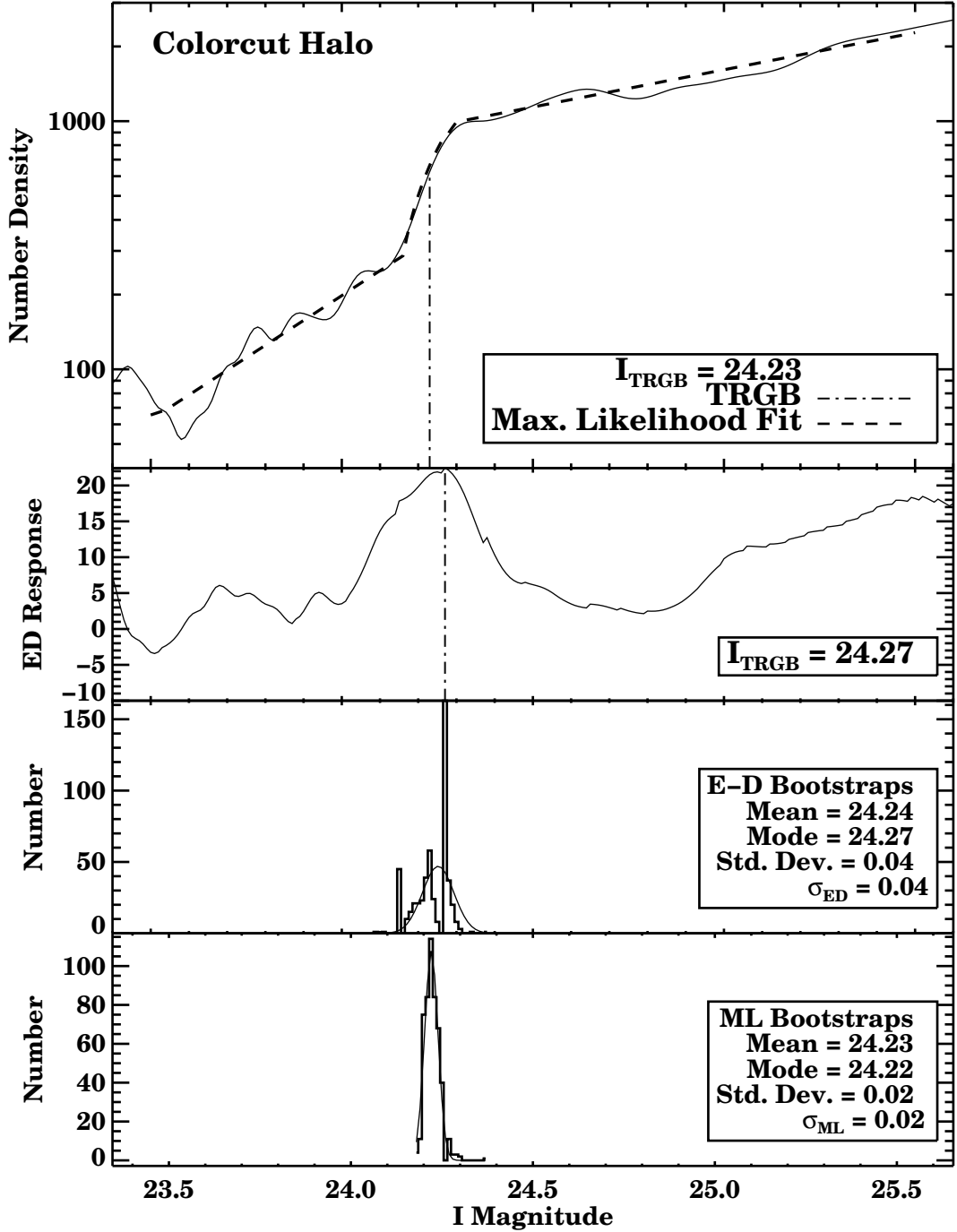}
\end{center}
\caption{Luminosity function of stars in the halo of UGC~06456 with $0.45 \leq (V-I) \leq 1.8$ mag. Overplotted is the best-fit model luminosity function determined via maximum likelihood analysis. The lower three panels are as in Fig.~\ref{figure:leoi_ccut_lf} }
\label{figure:ugc06456_halo_ccut_lf}
\end{figure}

The observations resolved the galaxy brilliantly into stars revealing a bright and knotted population of blue stars at its core and a smooth background of red stars. Fig.~\ref{figure:ugc06456_cmd} shows our color-magnitude diagram for the entire WFPC2 field. Evident in the CMD is a strong population of blue main sequence stars as well as a prominent red giant branch. There are also a small number of AGB stars visible above the red giant branch. We were able to remove the contamination from the blue stars by excluding the stars from the central region as shown in Fig.~\ref{figure:ugc06456_halo_reg}. Furthermore we exclude stars with colors outside the range $0.45 \leq (V-I) \leq 1.8$ mag. Fig.~\ref{figure:ugc06456_halo_ccut_lf} shows the resulting luminosity function.

\begin{figure}
\begin{center}
\includegraphics[height=8in]{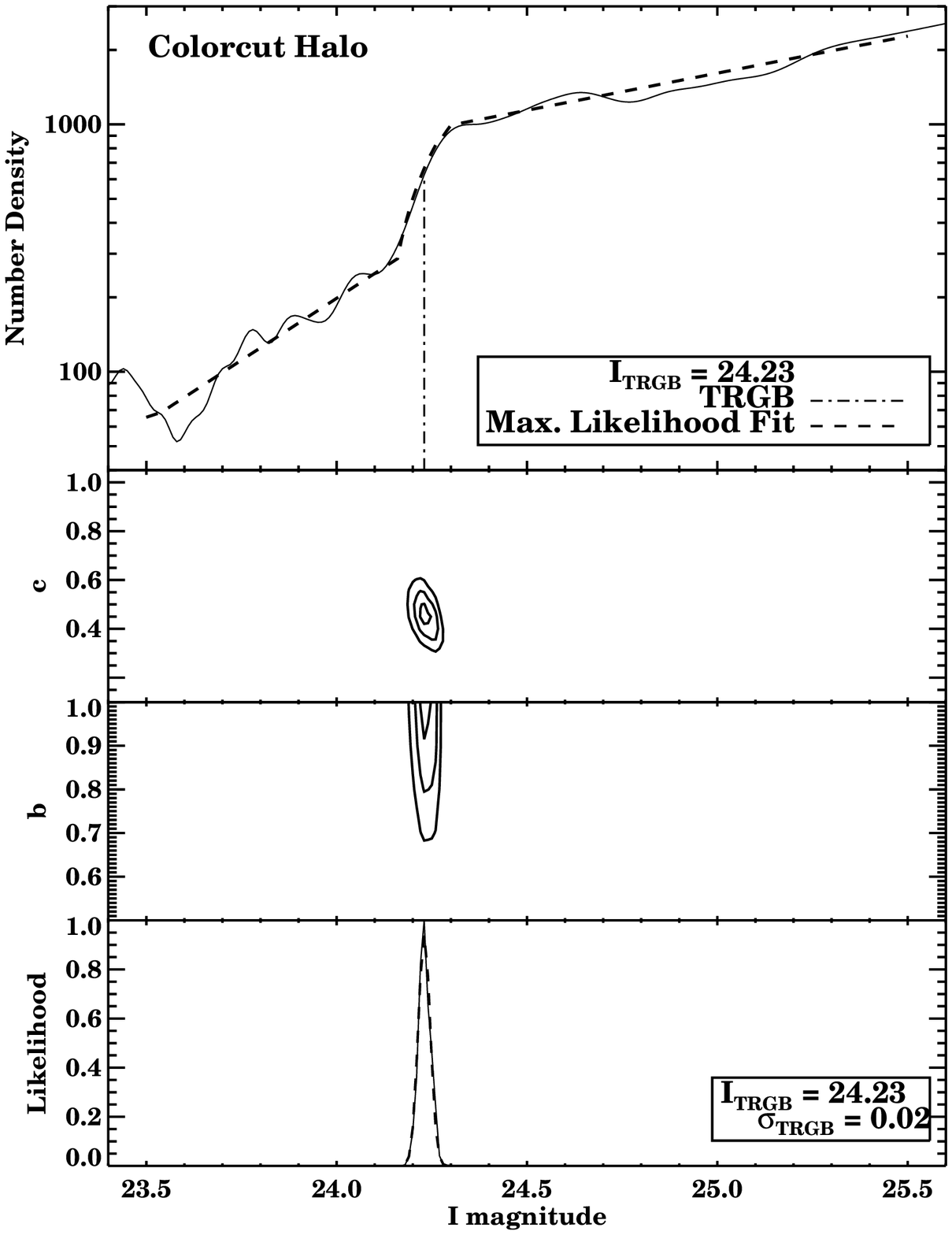}
\end{center}
\caption{Top panel as in Fig~\ref{figure:ugc06456_halo_ccut_lf}. The second and third panels plot $1\sigma$, $2\sigma$, and $3\sigma$ contours of the likelihood function parameters $c$ and $b$. The fourth panel plots the marginalized likelihood. A Gaussian fit with $\sigma_{\rm TRGB} = 0.02$ mag is overplotted as a dashed line. }
\label{figure:ugc06456_halo_ccut_like}
\end{figure}

 The TRGB is extremely obvious and the maximum likelihood analysis finds the TRGB at $I_{\rm TRGB} = 24.23 \pm 0.02$ mag, as shown in Fig.~\ref{figure:ugc06456_halo_ccut_like}, and the color of the TRGB is $(V-I)_{0 ,{{\rm TRGB}}} = 1.46 \pm 0.02$ mag. We thus take the metallicity of the red giant branch to be [Fe/H] = $-1.62 \pm 0.05$. Using an extinction of $A_I = 0.070$ mag we find the distance modulus for UGC~06456 to be $(m-M)_{\circ} = 28.19 \pm 0.04$(internal)$\pm 0.18$(systematic) mag and hence the heliocentric distance to be $4.34 \pm 0.07$ Mpc. This is in good agreement with the TRGB analysis carried out by previous authors using these same data \citep{lynds98, schulte98} who found $(m - M)_I = 28.25 \pm 0.10$ and $(m-M)_{\circ} = 28.4 \pm 0.09 \pm 0.18$ mag respectively.

\subsubsection{\bf UGC~03755}

UGC~03755 is a typical dwarf irregular galaxy that has been little studied by previous authors. It was of particular interest to us because it lies in the region of the sky with the strongest predicted infall (see Fig.~\ref{figure:skydist}).  

\begin{figure}
\begin{center}
\includegraphics{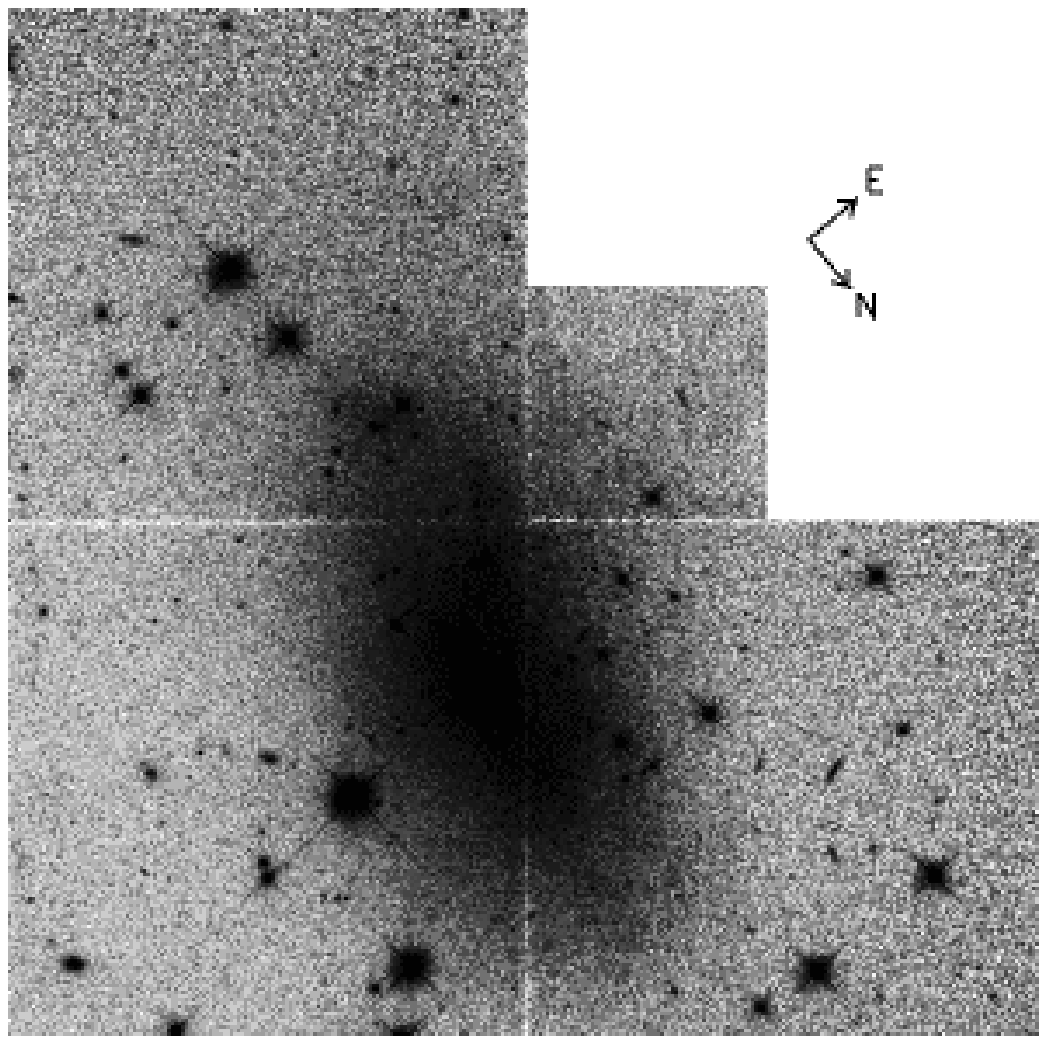}
\end{center}
\caption{Mosaic of a single 1300s {\em HST}/WFPC2 exposure of UGC~03755.}
\label{figure:ugc03755_image}
\end{figure}

\begin{figure}
\begin{center}
\includegraphics[height=8in]{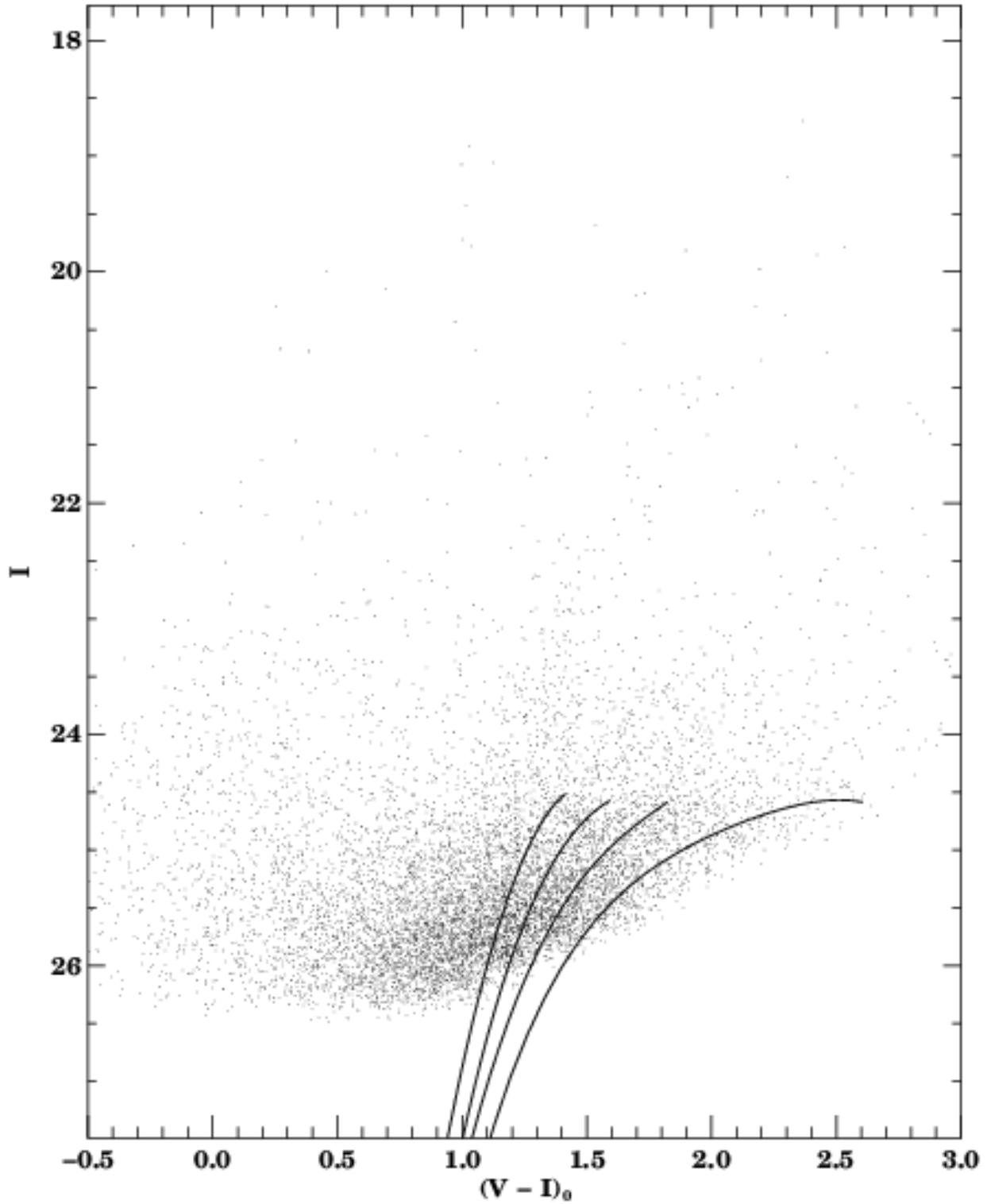}
\end{center}
\caption{Color--Magnitude Diagram for all stars in the UGC~03755 CCD frame. The red giant branch loci of M15, M2, NGC~1851, and 47 Tuc (from left to right; see Da Costa \& Armandroff 1990) are overplotted.}
\label{figure:ugc03755_cmd}
\end{figure}

\begin{figure}
\begin{center}
\includegraphics[height=8in]{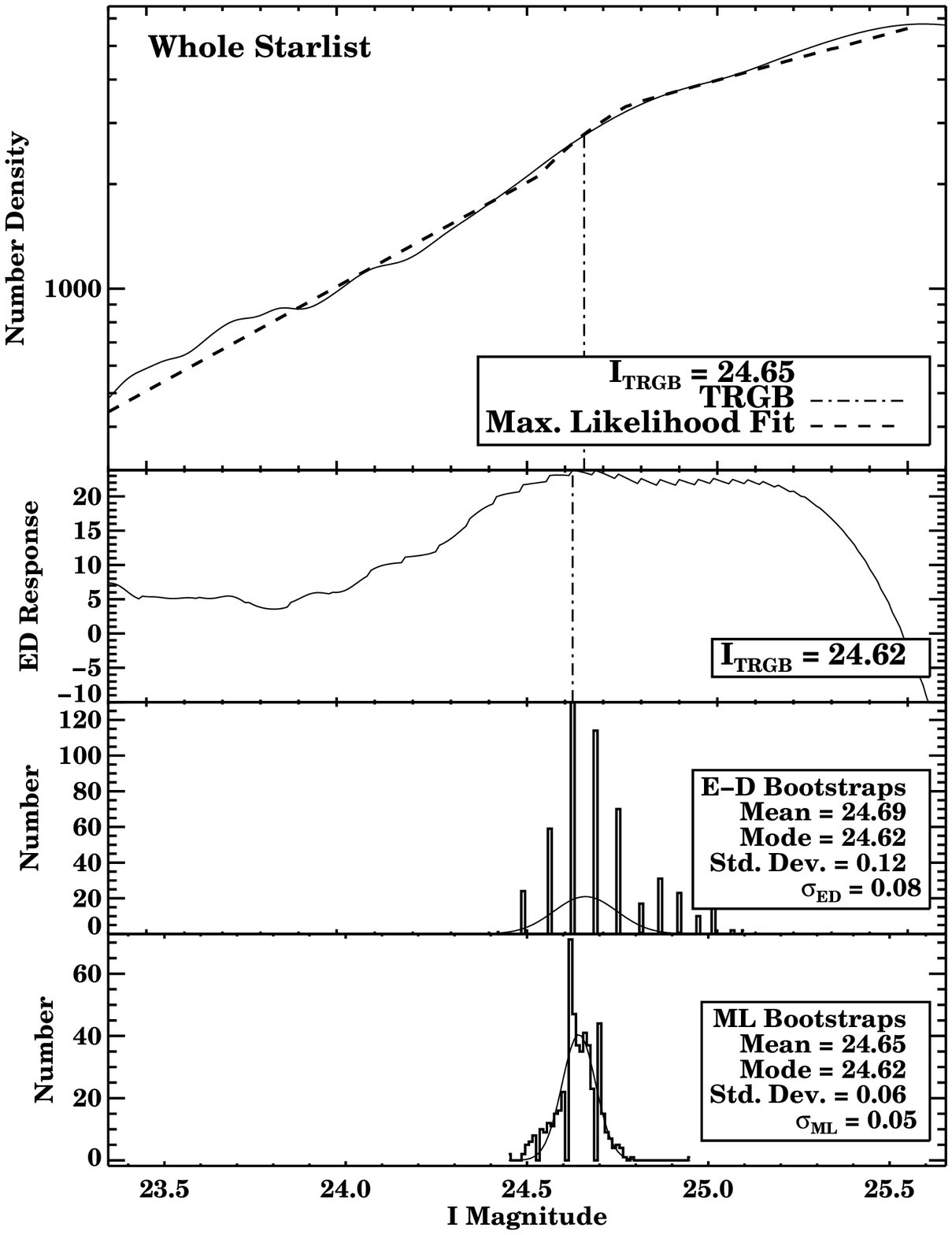}
\end{center}
\caption{Luminosity function of all stars in UGC~03755. Overplotted is the best-fit model luminosity function determined via maximum likelihood analysis. The lower three panels are as in Fig.~\ref{figure:leoi_ccut_lf} }
\label{figure:ugc03755_lf}
\end{figure}

\begin{figure}
\begin{center}
\includegraphics[height=8in]{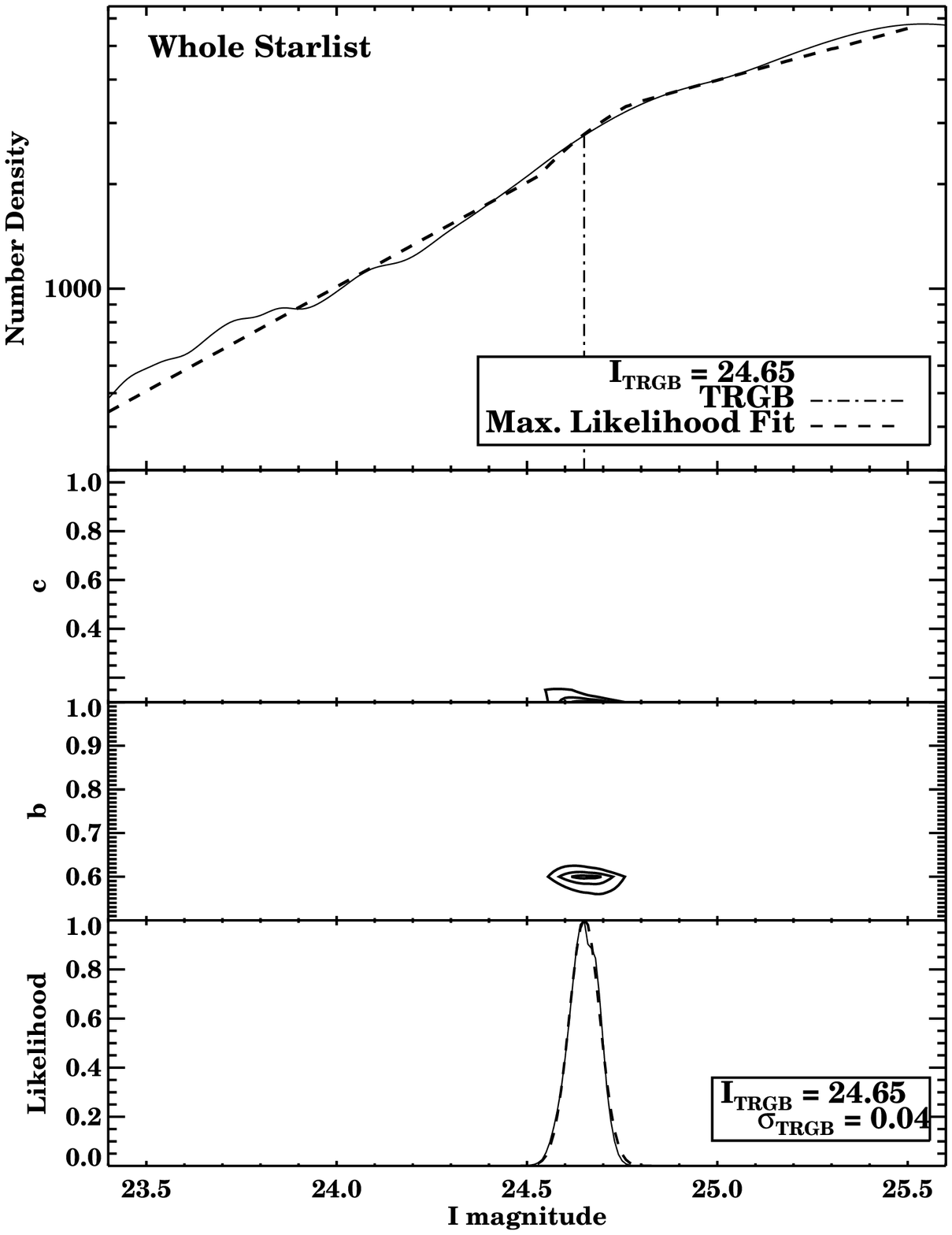}
\end{center}
\caption{Top panel as in Fig.~\ref{figure:ugc03755_lf}. The second and third panels plot $1\sigma$, $2\sigma$, and $3\sigma$ contours of the likelihood function parameters $c$ and $b$. The fourth panel plots the marginalized likelihood. A Gaussian fit with $\sigma_{\rm TRGB} = 0.04$ mag is overplotted as a dashed line.}
\label{figure:ugc03755_like}
\end{figure}

Fig.~\ref{figure:ugc03755_image} shows one of our F814W exposures of UGC~03755. The galaxy appears well resolved into stars. However, inspection of the CMD (Fig.~\ref{figure:ugc03755_cmd}) shows that we did not resolve the galaxy very deeply, especially in the F555W bandpass. There is only the slightest hint of a red giant branch, and the F555W photometry limits cut into it almost immediately. Making region or color cuts does not actually help a detection in this case. However, upon inspection of the luminosity function for UGC~03755 (Fig.~\ref{figure:ugc03755_lf}) we can see the red giant branch power-law signature for $\sim 0.8$ mag brighter than the magnitude limits in $I$ band. There is no strong discontinuity in the luminosity function as seen with the previous galaxies but there is a definite break in the power-law. We interpret this break as the tip of the red giant branch at $I_{\rm TRGB} = 24.65 \pm 0.06$ mag, as measured weakly by the maximum likelihood algorithm and shown in Fig.~\ref{figure:ugc03755_like}. Because the red giant branch is not well defined we chose to treat UGC~03755 as a case in which we did not have color data and hence we assume that its metallicity is within the calibrated range and adopt $M_{\rm I,TRGB} = -4.01 \pm 0.03$ mag. We use the \citet{schlegel98} line of sight extinction of $A_I = 0.172$ mag, which yields a distance modulus of $(m-M)_{\circ} = 28.49 \pm 0.07$(internal)$\pm 0.18$(systematic) mag. This corresponds to a heliocentric distance of $4.98 \pm 0.17$ Mpc. We take this as a lower limit on the distance to this galaxy, for if it were closer we should have seen a definitive red giant branch. Likewise we take this distance as the upper limit for our ability to measure distances in our {\em HST} dataset. Since all of our exposure times were the same we would not resolve the red giant branch population of any galaxy that is farther than this. 

\citet{georgiev97a} estimated $(m-M) = 28.08$ mag for this galaxy, based upon the magnitudes of the brightest resolved stars. In another paper \citep{georgiev97b} they estimated the calibration errors in their method to be $\pm 0.39$ mag. There is no strong agreement, but again we believe that UGC~03755 cannot be any closer than $\sim 5.0$ Mpc. 

\subsubsection{\bf NGC~2366 \& NGC~2683}

With success comes failure; we were not able to determine TRGB distances to many of our target galaxies. Based on our successful measurements and on work done by previous authors, we do not think that our failures were caused by contaminating stellar populations. Rather we think that in most cases the data were simply not deep enough to resolve out red giant branch stars. Hence, the galaxies are too far away for our data to reveal the TRGB. 

\begin{figure}
\begin{center}
\includegraphics[height=8in]{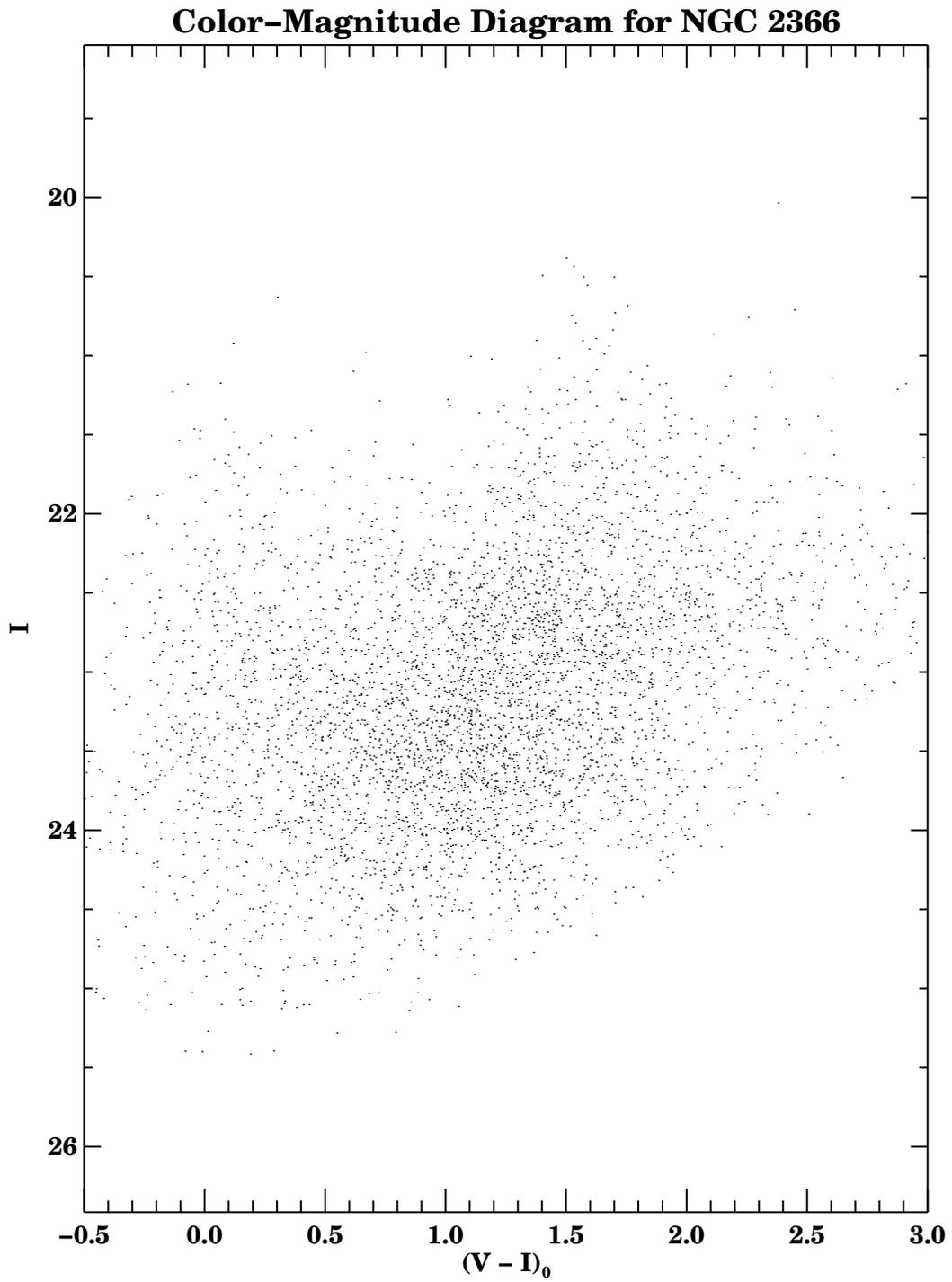}
\end{center}
\caption{NGC~2366 Color--Magnitude Diagram for all stars in the Keck/LRIS CCD frame.}
\label{figure:NGC2366_cmd}
\end{figure}

We estimate that in our Keck data we would not have been able to detect any red giant branch (or tip thereof) fainter than $I \sim 23.0$ mag. We take the limiting magnitude as at least $0.5$ magnitudes brighter than the turn over in the luminosity function. This means that we could not have successfully measured a TRGB distance for any galaxy with a distance greater than $2.5$ Mpc. This was unfortunate as many of our Keck targets are expected members of the M81 group. One example is NGC~2366. This dwarf irregular galaxy has been estimated to lie at a distance $d \leq 3.44$ Mpc, which is consistent with other M81 group members \citep[and references therein]{ferrarese00a}. Our CMD for this galaxy (Fig.~\ref{figure:NGC2366_cmd}) shows no recognizable structure, and the luminosity function turns over at about $I \sim 23.5$ mag never having any red giant branch power-law slope. Clearly our data for this object did not go deep enough to resolve a red giant branch population. 

\begin{figure}
\begin{center}
\includegraphics[height=8in]{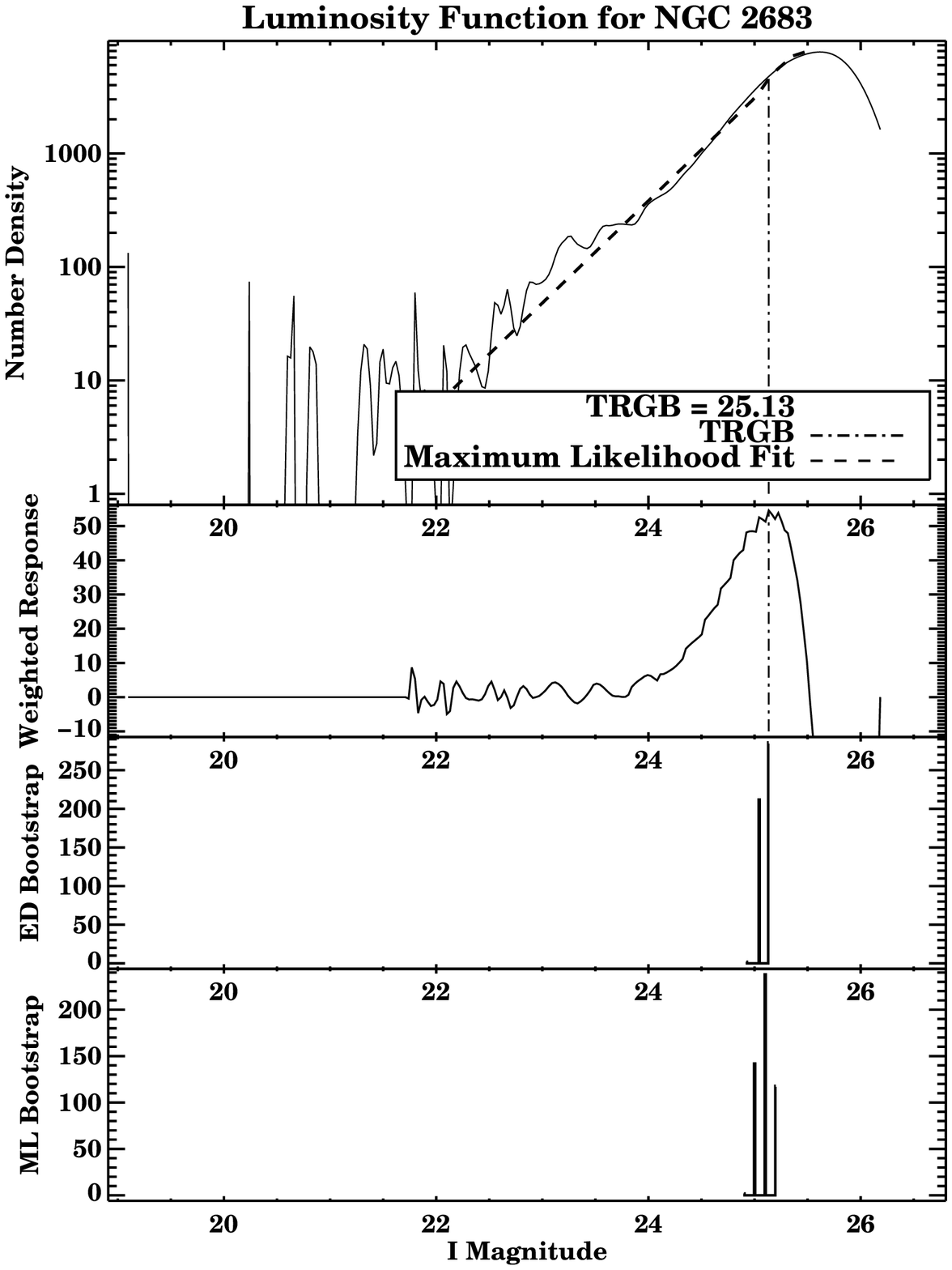}
\end{center}
\caption{Luminosity function of all stars in NGC~2683. Overplotted is the best-fit model luminosity function determined via maximum likelihood analysis. The lower three panels are as in Fig.~\ref{figure:leoi_ccut_lf} }
\label{figure:ngc2683_lf}
\end{figure}

The situation is much the same for our {\em HST} data. The typical detection limit in the {\em HST} data is $I \sim 25.0$ mag, corresponding to a linear distance limit of $6.2$ Mpc. An example of a galaxy we failed to measure a TRGB distance for is NGC~2683. This is a spiral galaxy seen nearly edge-on. We avoided the disk region and observed only in the halo region of the galaxy to minimize any Population I contamination. A recent survey of galaxies by \citet{tonry01} that used surface brightness fluctuations as a distance indicator gave a distance for this galaxy of $7.7$ Mpc. This galaxy is then clearly outside the reach of our data. There is little wonder that we see no hint of a red giant branch power-law in the luminosity function (Fig.~\ref{figure:ngc2683_lf}).  

\section{\bf Summary}

We have successfully used the TRGB as a distance indicator for seven local galaxies observed with {\em HST} and Keck. Using knowledge of the luminosity function behavior of the red giant branch, we conducted a maximum likelihood procedure to estimate the $I$ band magnitude of the TRGB. The uncertainty in the estimates of $I_{\rm TRGB}$ and $(V-I)_{\rm TRGB}$ were calculated via bootstrap resampling which makes no assumption about the underlying distributions and allows one to model the act of making repeated measurements of the same quantity. Bootstrap resampling also allows us to better estimate the error in $I_{\rm TRGB}$ when the number of red giant branch stars at that magnitude is small. We find that compared to Sobel edge detection the maximum likelihood analysis is equally accurate but more precise, and we find no systematic difference in the derived values of $I_{\rm TRGB}$ between the two methods. Indeed we find the weighted mean absolute difference in $I_{\rm TRGB}$ to be $<|I_{\rm TRGB,ML} - I_{\rm TRGB,ED}|> = 0.03 \pm 0.02$ mag (the simple average absolute difference in $I_{\rm TRGB}$ is $<|I_{\rm TRGB,ML} - I_{\rm TRGB,ED}|> = 0.03 \pm 0.01$ mag). The average ratio of uncertainties is $<\sigma_{\rm TRGB,ML} / \sigma_{\rm TRGB,ED}> = 0.62$. Therefore, on average, the maximum likelihood method is 38\% 
more precise than the Sobel edge detection. 
It should also be noted that other authors (e.g. \citet{cioni00}) have found Sobel edge detection to be a biased estimator of the TRGB. The bias depends on the amount of smoothing in the luminosity function at the location of the TRGB. For our data this bias was less than 0.04 mag which is less than random errors and therefore we would not be sensitive to it in a comparison to the maximum likelihood.
It is a positive feature of the TRGB method that it does not depend very sensitively on the statistical technique used to define the tip. The TRGB method will be most useful in further broadening our knowledge of distances to local galaxies and allow us to study the local flow in greater detail.

\acknowledgments

Bryan M\'{e}ndez acknowledges support from a Ford Foundation
Predoctoral Fellowship for Minorities. Some of the data presented
herein were obtained at the W.M. Keck Observatory, which is operated
as a scientific partnership among the California Institute of
Technology, the University of California and the National Aeronautics
and Space Administration. The Observatory was made possible by the
generous financial support of the W.M. Keck Foundation. Support for
proposal \#~8199 was provided by NASA through a grant from the Space
Telescope Science Institute, which is operated by the Association of
Universities for Research in Astronomy, Inc., under NASA contract NAS
5-26555. This research  has made use  of the NASA/IPAC  Extragalactic
Database (NED) which  is operated by the Jet  Propulsion Laboratory,
California Institute of Technology, under  contract with the National
Aeronautics and Space Administration.

\bibliographystyle{apj}
\parskip=0pt
\begin{small}

\begin{thebibliography}{}

\bibitem[\protect\citeauthoryear{{Babu} \& {Feigelson}}{{Babu} \&
  {Feigelson}}{1996}]{babu96}
{Babu}, G.~J.,  \& {Feigelson}, E.~D., ed. 1996, {Astrostatistics
  interdisciplinary statistics}

\bibitem[\protect\citeauthoryear{{Baker} et~al.}{{Baker}
  et~al.}{1998}]{baker98}
{Baker}, J.~E., {Davis}, M., {Strauss}, M.~A., {Lahav}, O.,  \& {Santiago},
  B.~.~X. 1998, \apj, 508, 6

\bibitem[\protect\citeauthoryear{{Beland}, {Boulade}, \& {Davidge}}{{Beland}
  et~al.}{1988}]{beland88}
{Beland}, S., {Boulade}, O.,  \& {Davidge}, T. 1988, CFHT Info. Bull., 19, 16

\bibitem[\protect\citeauthoryear{{Bellazzini}, {Ferraro}, \&
  {Pancino}}{{Bellazzini} et~al.}{2001}]{bellazzini01}
{Bellazzini}, M., {Ferraro}, F.~R.,  \& {Pancino}, E. 2001, \apj, 556, 635

\bibitem[\protect\citeauthoryear{{Burrows}}{{Burrows}}{1994}]{burrows94}
{Burrows}, C.~J., ed. 1994, {Hubble Space Telescope Wide Field and Planetary
  Camera 2 Instrument handbook, version 2.0}

\bibitem[\protect\citeauthoryear{{Burstein}}{{Burstein}}{1990}]{burstein90}
{Burstein}, D. 1990, Reports on Progress in Physics, 53, 421

\bibitem[\protect\citeauthoryear{{Caputo} et~al.}{{Caputo}
  et~al.}{1999}]{caputo99}
{Caputo}, F., {Cassisi}, S., {Castellani}, M., {Marconi}, G.,  \&
  {Santolamazza}, P. 1999, \aj, 117, 2199

\bibitem[\protect\citeauthoryear{{Cioni} et~al.}{{Cioni}
  et~al.}{2000}]{cioni00}
{Cioni}, M.-R.~L., {van der Marel}, R.~P., {Loup}, C.,  \& {Habing}, H.~J.
  2000, \aap, 359, 601

\bibitem[\protect\citeauthoryear{{Da Costa} \& {Armandroff}}{{Da Costa} \&
  {Armandroff}}{1990}]{dacosta90}
{Da Costa}, G.~S.,  \& {Armandroff}, T.~E. 1990, \aj, 100, 162

\bibitem[\protect\citeauthoryear{{de Vaucouleurs}}{{de
  Vaucouleurs}}{1963}]{devaucouleurs63}
{de Vaucouleurs}, G. 1963, \apj, 137, 720

\bibitem[\protect\citeauthoryear{{Dolphin}}{{Dolphin}}{2000a}]{dolphin00b}
{Dolphin}, A.~E. 2000a, \pasp, 112, 1397

\bibitem[\protect\citeauthoryear{{Dolphin}}{{Dolphin}}{2000b}]{dolphin00a}
{Dolphin}, A.~E. 2000b, \pasp, 112, 1383

\bibitem[\protect\citeauthoryear{{Dolphin} et~al.}{{Dolphin}
  et~al.}{2001}]{dolphin01}
{Dolphin}, A.~E., et~al. 2001, \mnras, 324, 249

\bibitem[\protect\citeauthoryear{{Ferrarese} et~al.}{{Ferrarese}
  et~al.}{2000a}]{ferrarese00a}
{Ferrarese}, L., et~al. 2000a, \apjs, 128, 431

\bibitem[\protect\citeauthoryear{{Ferrarese} et~al.}{{Ferrarese}
  et~al.}{2000b}]{ferrarese00b}
{Ferrarese}, L., et~al. 2000b, \apj, 529, 745

\bibitem[\protect\citeauthoryear{{Fisher} et~al.}{{Fisher}
  et~al.}{1994}]{fisher94}
{Fisher}, K.~B., {Davis}, M., {Strauss}, M.~A., {Yahil}, A.,  \& {Huchra},
  J.~P. 1994, \mnras, 267, 927

\bibitem[\protect\citeauthoryear{{Georgiev}, {Bilkina}, \&
  {Dencheva}}{{Georgiev} et~al.}{1997}]{georgiev97b}
{Georgiev}, T.~B., {Bilkina}, B.~I.,  \& {Dencheva}, N.~M. 1997, Astronomy
  Letters, 23, 656

\bibitem[\protect\citeauthoryear{{Georgiev}, {Karachentsev}, \&
  {Tikhonov}}{{Georgiev} et~al.}{1997}]{georgiev97a}
{Georgiev}, T.~B., {Karachentsev}, I.~D.,  \& {Tikhonov}, N.~A. 1997, Astronomy
  Letters, 23, 514

\bibitem[\protect\citeauthoryear{{Holtzman} et~al.}{{Holtzman}
  et~al.}{1995}]{holtzman95}
{Holtzman}, J.~A., et~al. 1995, \pasp, 107, 156

\bibitem[\protect\citeauthoryear{{Iben} \& {Renzini}}{{Iben} \&
  {Renzini}}{1983}]{iben83}
{Iben}, I.,  \& {Renzini}, A. 1983, \araa, 21, 271

\bibitem[\protect\citeauthoryear{{Karachentsev} \& {Makarov}}{{Karachentsev} \&
  {Makarov}}{1996}]{karachentsev96}
{Karachentsev}, I.~D.,  \& {Makarov}, D.~A. 1996, \aj, 111, 794

\bibitem[\protect\citeauthoryear{{Karachentsev} \& {Tikhonov}}{{Karachentsev}
  \& {Tikhonov}}{1994}]{karachentsev94}
{Karachentsev}, I.~D.,  \& {Tikhonov}, N.~A. 1994, \aap, 286, 718

\bibitem[\protect\citeauthoryear{{Kim} et~al.}{{Kim} et~al.}{2002}]{kim02}
{Kim}, M., {Kim}, E., {Lee}, M.~G., {Sarajedini}, A.,  \& {Geisler}, D. 2002,
  \aj, 123, 244

\bibitem[\protect\citeauthoryear{{Krist} \& {Hook}}{{Krist} \&
  {Hook}}{1996}]{krist96}
{Krist}, J.,  \& {Hook}, R., ed. 1996, {Tiny Tim User's Manual, V4.2}

\bibitem[\protect\citeauthoryear{{Landolt}}{{Landolt}}{1992}]{landolt92}
{Landolt}, A.~U. 1992, \aj, 104, 340

\bibitem[\protect\citeauthoryear{{Lee} et~al.}{{Lee} et~al.}{1993}]{lee93b}
{Lee}, M.~G., {Freedman}, W., {Mateo}, M., {Thompson}, I., {Roth}, M.,  \&
  {Ruiz}, M. 1993, \aj, 106, 1420

\bibitem[\protect\citeauthoryear{{Lee}, {Freedman}, \& {Madore}}{{Lee}
  et~al.}{1993}]{lee93}
{Lee}, M.~G., {Freedman}, W.~L.,  \& {Madore}, B.~F. 1993, \apj, 417, 553

\bibitem[\protect\citeauthoryear{{Lynds} et~al.}{{Lynds}
  et~al.}{1998}]{lynds98}
{Lynds}, R., {Tolstoy}, E., {O'Neil.}, E.~J.,  \& {Hunter}, D.~A. 1998, \aj,
  116, 146

\bibitem[\protect\citeauthoryear{{Lyo} \& {Lee}}{{Lyo} \& {Lee}}{1997}]{lyo97}
{Lyo}, A.-R.,  \& {Lee}, M.~G. 1997, Journal of Korean Astronomical Society,
  30, 27

\bibitem[\protect\citeauthoryear{{Madore} \& {Freedman}}{{Madore} \&
  {Freedman}}{1995}]{madore95}
{Madore}, B.~F.,  \& {Freedman}, W.~L. 1995, \aj, 109, 1645

\bibitem[\protect\citeauthoryear{{Madore} \& {Freedman}}{{Madore} \&
  {Freedman}}{1998}]{madore98}
{Madore}, B.~F.,  \& {Freedman}, W.~L. 1998, in Stellar astrophysics for the
  local group: VIII Canary Islands Winter School of Astrophysics, 263

\bibitem[\protect\citeauthoryear{{M\'{e}ndez} et~al.}{{M\'{e}ndez}
  et~al.}{2002}]{mendez02b}
{M\'{e}ndez}, B., {Davis}, M., {Newman}, J., {Madore}, B.~F., {Freedman},
  W.~L.,  \& {Moustakas}, J. 2002, {In Preparation}

\bibitem[\protect\citeauthoryear{{Minniti}, {Zijlstra}, \& {Alonso}}{{Minniti}
  et~al.}{1999}]{minniti99}
{Minniti}, D., {Zijlstra}, A.~A.,  \& {Alonso}, M.~V. 1999, \aj, 117, 881

\bibitem[\protect\citeauthoryear{{Mould} \& {Kristian}}{{Mould} \&
  {Kristian}}{1986}]{mould86}
{Mould}, J.,  \& {Kristian}, J. 1986, \apj, 305, 591

\bibitem[\protect\citeauthoryear{{Nusser} \& {Davis}}{{Nusser} \&
  {Davis}}{1994}]{nusser94}
{Nusser}, A.,  \& {Davis}, M. 1994, \apjl, 421, L1

\bibitem[\protect\citeauthoryear{{Oke} et~al.}{{Oke} et~al.}{1995}]{oke95}
{Oke}, J.~B., et~al. 1995, \pasp, 107, 375

\bibitem[\protect\citeauthoryear{{Peebles}}{{Peebles}}{1993}]{peebles93}
{Peebles}, P.~J.~E. 1993, {Principles of physical cosmology} (Princeton Series
  in Physics, Princeton, NJ: Princeton University Press, |c1993)

\bibitem[\protect\citeauthoryear{{Renzini} \& {Fusi Pecci}}{{Renzini} \& {Fusi
  Pecci}}{1988}]{renzini88}
{Renzini}, A.,  \& {Fusi Pecci}, F. 1988, \araa, 26, 199

\bibitem[\protect\citeauthoryear{{Rozanski} \& {Rowan-Robinson}}{{Rozanski} \&
  {Rowan-Robinson}}{1994}]{rozanski94}
{Rozanski}, R.,  \& {Rowan-Robinson}, M. 1994, \mnras, 271, 530

\bibitem[\protect\citeauthoryear{{Saha} et~al.}{{Saha} et~al.}{2001}]{saha01}
{Saha}, A., {Sandage}, A., {Tammann}, G.~A., {Dolphin}, A.~E., {Christensen},
  J., {Panagia}, N.,  \& {Macchetto}, F.~D. 2001, \apj, 562, 314

\bibitem[\protect\citeauthoryear{{Sakai} \& {Madore}}{{Sakai} \&
  {Madore}}{1999}]{sakai99}
{Sakai}, S.,  \& {Madore}, B.~F. 1999, \apj, 526, 599

\bibitem[\protect\citeauthoryear{{Sakai}, {Madore}, \& {Freedman}}{{Sakai}
  et~al.}{1996}]{sakai96}
{Sakai}, S., {Madore}, B.~F.,  \& {Freedman}, W.~L. 1996, \apj, 461, 713

\bibitem[\protect\citeauthoryear{{Sakai}, {Madore}, \& {Freedman}}{{Sakai}
  et~al.}{1997}]{sakai97}
{Sakai}, S., {Madore}, B.~F.,  \& {Freedman}, W.~L. 1997, \apj, 480, 589

\bibitem[\protect\citeauthoryear{{Sakai}, {Zaritsky}, \& {Kennicutt}}{{Sakai}
  et~al.}{2000}]{sakai00}
{Sakai}, S., {Zaritsky}, D.,  \& {Kennicutt}, R.~C. 2000, \aj, 119, 1197

\bibitem[\protect\citeauthoryear{{Salaris} \& {Cassisi}}{{Salaris} \&
  {Cassisi}}{1997}]{salaris97}
{Salaris}, M.,  \& {Cassisi}, S. 1997, \mnras, 289, 406

\bibitem[\protect\citeauthoryear{{Sandage}, {Tammann}, \& {Yahil}}{{Sandage}
  et~al.}{1979}]{sandage79}
{Sandage}, A., {Tammann}, G.~A.,  \& {Yahil}, A. 1979, \apj, 232, 352

\bibitem[\protect\citeauthoryear{{Schechter}, {Mateo}, \& {Saha}}{{Schechter}
  et~al.}{1993}]{schechter93}
{Schechter}, P.~L., {Mateo}, M.,  \& {Saha}, A. 1993, \pasp, 105, 1342

\bibitem[\protect\citeauthoryear{{Schlegel}, {Finkbeiner}, \&
  {Davis}}{{Schlegel} et~al.}{1998}]{schlegel98}
{Schlegel}, D.~J., {Finkbeiner}, D.~P.,  \& {Davis}, M. 1998, \apj, 500, 525

\bibitem[\protect\citeauthoryear{{Schulte-Ladbeck}, {Crone}, \&
  {Hopp}}{{Schulte-Ladbeck} et~al.}{1998}]{schulte98}
{Schulte-Ladbeck}, R.~E., {Crone}, M.~M.,  \& {Hopp}, U. 1998, \apjl, 493, L23

\bibitem[\protect\citeauthoryear{{Seitzer} et~al.}{{Seitzer}
  et~al.}{2001}]{seitzer01}
{Seitzer}, P., et~al. 2001, in American Astronomical Society Meeting, Vol.~33,
  800

\bibitem[\protect\citeauthoryear{{Shaklan}, {Sharman}, \& {Pravdo}}{{Shaklan}
  et~al.}{1995}]{shaklan95}
{Shaklan}, S., {Sharman}, M.~C.,  \& {Pravdo}, S.~H. 1995, Appl. Opt., 34, 6672

\bibitem[\protect\citeauthoryear{{Stetson}}{{Stetson}}{1987}]{stetson87}
{Stetson}, P.~B. 1987, \pasp, 99, 191

\bibitem[\protect\citeauthoryear{{Stetson}}{{Stetson}}{1994}]{stetson94}
{Stetson}, P.~B. 1994, \pasp, 106, 250

\bibitem[\protect\citeauthoryear{{Stetson}}{{Stetson}}{2000}]{stetson00}
{Stetson}, P.~B. 2000, \pasp, 112, 925

\bibitem[\protect\citeauthoryear{{Tikhonov} \& {Karachentsev}}{{Tikhonov} \&
  {Karachentsev}}{1998}]{tikhonov98}
{Tikhonov}, N.~A.,  \& {Karachentsev}, I.~D. 1998, \aaps, 128, 325

\bibitem[\protect\citeauthoryear{{Tonry} et~al.}{{Tonry}
  et~al.}{2001}]{tonry01}
{Tonry}, J.~L., {Dressler}, A., {Blakeslee}, J.~P., {Ajhar}, E.~A., {Fletcher},
  A.~., {Luppino}, G.~A., {Metzger}, M.~R.,  \& {Moore}, C.~B. 2001, \apj, 546,
  681

\bibitem[\protect\citeauthoryear{{Tully}}{{Tully}}{1988}]{tully88}
{Tully}, R.~B. 1988, {Nearby galaxies catalog} (Cambridge and New York,
  Cambridge University Press, 1988, 221 p.)

\bibitem[\protect\citeauthoryear{{Udalski} et~al.}{{Udalski}
  et~al.}{2001}]{udalski01}
{Udalski}, A., {Wyrzykowski}, L., {Pietrzynski}, G., {Szewczyk}, O.,
  {Szymanski}, M., {Kubiak}, M., {Soszynski}, I.,  \& {Zebrun}, K. 2001, Acta
  Astronomica, 51, 221

\bibitem[\protect\citeauthoryear{{Wyder}}{{Wyder}}{2001}]{wyder01}
{Wyder}, T.~K. 2001, \aj, 122, 2490

\bibitem[\protect\citeauthoryear{{Zoccali} \& {Piotto}}{{Zoccali} \&
  {Piotto}}{2000}]{zoccali00}
{Zoccali}, M.,  \& {Piotto}, G. 2000, \aap, 358, 943

\end{thebibliography}

\end{small}

\clearpage

\clearpage

\pagebreak

\begin{deluxetable}{lrrcllrrrl}
\rotate
\tabletypesize{\scriptsize}
\tablenum{1}
\tablecolumns{10}
\tablewidth{0pt}
\tablecaption{Details of Observations}
\tablehead{
\colhead{Galaxy} & \colhead{l} & \colhead{b} & \colhead{Date} & \colhead{Instrument} & \colhead{Filter} & \colhead{\# Exp.} & \colhead{Exp. Time/Frame} & \colhead{FWHM} & \colhead{Comments} \\
\colhead{} & \colhead{($^\circ$)} & \colhead{($^\circ$)} & \colhead{(dd/mm/yy)} & \colhead{} & \colhead{} & \colhead{} & \colhead{(sec)} & \colhead{(arcsec)} & \colhead{} \\
}
\startdata
Leo I           & 225.982 & 49.110 & 23/12/97 & Keck/LRIS & I/V         & 2/1 & 300/300   & 0.85-0.94/0.92   & Success \\
Sextans B       & 233.200 & 43.784 & 22/12/97 & Keck/LRIS & I/V         & 1/1 & 300/300   & 0.75/0.78        & Success \\
Holmberg I      & 140.729 & 38.662 & 23/12/97 & Keck/LRIS & I/V         & 1/1 & 300/300   & 0.97/1.08        & Poor Image \\
Holmberg II     & 144.283 & 32.691 & 22/12/97 & Keck/LRIS & I/V         & 4/1 & 300/600   & 0.82-0.93/0.86   & Shallow \\ 
Holmberg IX     & 141.977 & 41.052 & 22/12/97 & Keck/LRIS & I/V         & 1/1 & 300/300   & 0.75/0.88        & Shallow \\
IC 342          & 138.173 & 10.580 & 22/12/97 & Keck/LRIS & I/V         & 4/1 & 300/600   & 0.77-0.90/1.02   & Crowded \\ 
IC 2574         & 140.210 & 43.605 & 23/12/97 & Keck/LRIS & I/V         & 3/1 & 250-300/300 & 0.91-1.02/1.02 & Shallow \\
NGC~1560        & 138.367 & 16.022 & 22/12/97 & Keck/LRIS & I/V         & 4/1 & 300/600   & 1.10-1.29/1.40   & Shallow \\
NGC~2366        & 146.419 & 28.535 & 22/12/97 & Keck/LRIS & I/V         & 4/1 & 300/600   & 0.77-1.11/0.94   & Shallow \\
NGC~2903        & 208.711 & 44.540 & 22/12/97 & Keck/LRIS & I/V         & 3/1 & 300/400   & 0.74-0.78/0.86   & Shallow \\
NGC~2976        & 143.914 & 40.902 & 22/12/97 & Keck/LRIS & I/V         & 3/1 & 300/400   & 0.87-0.94/0.84   & Shallow \\
NGC~3109        & 262.101 & 23.070 & 22/12/97 & Keck/LRIS & I/V         & 1/1 & 300/300   & 0.88/0.90        & Success \\
UGC~02684       & 166.322 &-32.747 & 22/12/97 & Keck/LRIS & I/V         & 10/2& 300-400/600-800& 0.61-0.89/0.72-0.96 & Unresolved \\
UGC~02716       & 166.777 &-31.825 & 22/12/97 & Keck/LRIS & I/V         & 7/2 & 300-400/600-800&  $\cdots$   & Unresolved \\
UGC~03974       & 203.100 & 18.541 & 22/12/97 & Keck/LRIS & I/V         & 8/2 & 400-500/800-900&  $\cdots$   & Unresolved \\
UGC~04115       & 207.003 & 20.897 & 22/12/97 & Keck/LRIS & I/V         & 8/2 & 300-400/600-800& 0.64-0.83/0.76 & Unresolved \\
UGC~05272       & 195.416 & 50.563 & 23/12/97 & Keck/LRIS & I/V         & 4/1 & 300/600   & 0.77-0.80/0.78   & Unresolved \\
UGC~05340       & 199.887 & 51.614 & 23/12/97 & Keck/LRIS & I/V         & 4/1 & 300/600   & 0.74-0.92/0.94   & Unresolved \\
IC 342          & 138.173 & 10.580 & 14/08/99 & {\em HST}/WFPC2 & F814W/F555W & 2/2 & 1300/1300 &    $\cdots$      & Crowded \\
IRAS 12483-1311 & 302.747 & 49.415 & 22/01/00 & {\em HST}/WFPC2 & F814W       & 2   & 1200      &    $\cdots$      & Bad z \\
NGC~1313        & 283.472 &-44.654 & 14/01/00 & {\em HST}/WFPC2 & F814W       & 2   & 1300      &    $\cdots$      & Success \\
NGC~2683        & 190.457 & 38.761 & 15/02/00 & {\em HST}/WFPC2 & F814W       & 2   & 1300      &    $\cdots$      & Shallow \\
NGC~2903        & 208.711 & 44.540 & 02/05/00 & {\em HST}/WFPC2 & F814W       & 2   & 1200      &    $\cdots$      & Shallow \\
NGC~6503        & 100.574 & 30.640 & 08/11/99 & {\em HST}/WFPC2 & F814W       & 2   & 1300      &    $\cdots$      & Shallow \\
UGC~03476       & 180.766 & 10.513 & 13/02/00 & {\em HST}/WFPC2 & F814W/F555W & 2/2 & 1300/1300 &    $\cdots$      & Shallow \\
UGC~03698       & 172.971 & 21.622 & 11/02/00 & {\em HST}/WFPC2 & F814W       & 2   & 1300      &    $\cdots$      & Shallow \\
UGC~03755       & 206.013 &  9.715 & 26/03/00 & {\em HST}/WFPC2 & F814W/F555W & 2/2 & 1200/1200 &    $\cdots$      & Marginal \\
UGC~03860       & 177.806 & 23.934 & 02/07/00 & {\em HST}/WFPC2 & F814W       & 2   & 1300      &    $\cdots$      & Shallow \\
UGC~03974       & 203.100 & 18.541 & 22/01/00 & {\em HST}/WFPC2 & F814W       & 2   & 1200      &    $\cdots$      & Shallow \\
UGC~04115       & 207.003 & 20.897 & 24/09/99 & {\em HST}/WFPC2 & F814W       & 2   & 1200      &    $\cdots$      & Shallow \\
UGC~07577       & 137.751 & 72.945 & 09/05/00 & {\em HST}/WFPC2 & F814W       & 2   & 1300      &    $\cdots$      & Success \\
UGC~06456$^{a}$	& 127.836 & 37.327 & 10/07/95 & {\em HST}/WFPC2 & F814W/F555W  & 3/3 & 1400/1400 &   $\cdots$      & Success \\

\enddata
\tablenotetext{a}{{\em HST}/WFPC2 archival object.}
\label{table:obs}
\end{deluxetable}

\clearpage
\begin{deluxetable}{llllllllll}
\rotate
\tabletypesize{\scriptsize}
\tablenum{2}
\tablecolumns{10}
\tablewidth{0pt}
\tablecaption{Distances to Nearby Galaxies}
\tablehead{\colhead{Galaxy} & \colhead{$I_{\rm TRGB}$} & \colhead{$(V-I)_{\circ ,{\rm TRGB}}$} & \colhead{$(V-I)_{-3.5}$} & \colhead{[Fe/H]} & \colhead{$M_{I, {\rm TRGB}}$} & \colhead{$A_V$} & \colhead{$A_I$} & \colhead{$\mu_{\circ}$} & \colhead{d$_{\odot}$}  \\ \colhead{} & \colhead{(mag $\pm$ r $\pm$ s)} & \colhead{(mag)} & \colhead{(mag)} & \colhead{(dex)} & \colhead{(mag $\pm$ r $\pm$ s)} & \colhead{(mag)} & \colhead{(mag)} & \colhead{(mag)} & \colhead{(Mpc)} }
\startdata
\cutinhead{Using Maximum Likelihood Method}
Leo I & $18.14 \pm 0.07 \pm 0.05$ & $1.39 \pm 0.02$ & $1.28 \pm 0.02$ & $-1.94 \pm 0.08$ & $-3.98 \pm 0.02 \pm 0.04$ & $0.12 \pm 0.01$ & $0.07 \pm 0.01$ & $22.05 \pm 0.10$ & $0.26 \pm 0.01$ \cr
NGC 3109 & $21.63 \pm 0.05 \pm 0.03$ & $1.46 \pm 0.02$ & $1.34 \pm 0.02$ & $-1.69 \pm 0.06$ & $-4.02 \pm 0.01 \pm 0.02$ & $0.22 \pm 0.02$ & $0.13 \pm 0.01$ & $25.52 \pm 0.06$ & $1.27 \pm 0.04$ \cr
Sextans B & $21.68 \pm 0.02 \pm 0.02$ & $1.44 \pm 0.01$ & $1.33 \pm 0.01$ & $-1.73 \pm 0.04$ & $-4.01 \pm 0.01 \pm 0.02$ & $0.10 \pm 0.01$ & $0.06 \pm 0.01$ & $25.63 \pm 0.04$ & $1.34 \pm 0.02$ \cr
UGC 07577 & $23.01 \pm 0.06 \pm 0.02$ & $\cdots$ & $\cdots$ & $\cdots$ & $-4.01 \pm 0.03$ & $0.07 \pm 0.01$ & $0.04 \pm 0.00$ & $26.98 \pm 0.07$ & $2.49 \pm 0.08$ \cr
NGC 1313 & $24.28 \pm 0.04 \pm 0.02$ & $\cdots$ & $\cdots$ & $\cdots$ & $-4.01 \pm 0.03$ & $0.36 \pm 0.04$ & $0.21 \pm 0.02$ & $28.08 \pm 0.06$ & $4.13 \pm 0.11$ \cr
UGC 06456 & $24.23 \pm 0.02 \pm 0.02$ & $1.46 \pm 0.02$ & $1.36 \pm 0.01$ & $-1.62 \pm 0.05$ & $-4.03 \pm 0.01 \pm 0.01$ & $0.12 \pm 0.01$ & $0.07 \pm 0.01$ & $28.19 \pm 0.04$ & $4.34 \pm 0.07$ \cr
UGC 03755 & $24.65 \pm 0.06 \pm 0.02$ & $\cdots$ & $\cdots$ & $\cdots$ & $-4.01 \pm 0.03$ & $0.29 \pm 0.03$ & $0.17 \pm 0.02$ & $28.49 \pm 0.07$ & $4.98 \pm 0.17$ \cr
\cutinhead{Using Edge-Detection Method}
Leo I & $18.14 \pm 0.11 \pm 0.05$ & $1.39 \pm 0.02$ & $1.28 \pm 0.02$ & $-1.93 \pm 0.07$ & $-3.99 \pm 0.01 \pm 0.04$ & $0.12 \pm 0.01$ & $0.07 \pm 0.01$ & $22.06 \pm 0.13$ & $0.26 \pm 0.02$ \cr
NGC 3109 & $21.60 \pm 0.08 \pm 0.03$ & $1.46 \pm 0.02$ & $1.35 \pm 0.02$ & $-1.66 \pm 0.06$ & $-4.02 \pm 0.01 \pm 0.02$ & $0.22 \pm 0.02$ & $0.13 \pm 0.01$ & $25.49 \pm 0.09$ & $1.25 \pm 0.05$ \cr
Sextans B & $21.68 \pm 0.05 \pm 0.02$ & $1.44 \pm 0.01$ & $1.33 \pm 0.01$ & $-1.73 \pm 0.03$ & $-4.01 \pm 0.01 \pm 0.02$ & $0.10 \pm 0.01$ & $0.06 \pm 0.01$ & $25.63 \pm 0.06$ & $1.34 \pm 0.04$ \cr
UGC 07577 & $23.01 \pm 0.09 \pm 0.02$ & $\cdots$ & $\cdots$ & $\cdots$ & $-4.01 \pm 0.03$ & $0.07 \pm 0.01$ & $0.04 \pm 0.00$ & $26.98 \pm 0.10$ & $2.49 \pm 0.11$ \cr
NGC 1313 & $24.35 \pm 0.05 \pm 0.02$ & $\cdots$ & $\cdots$ & $\cdots$ & $-4.01 \pm 0.03$ & $0.36 \pm 0.04$ & $0.21 \pm 0.02$ & $28.15 \pm 0.07$ & $4.26 \pm 0.13$ \cr
UGC 06456 & $24.27 \pm 0.04 \pm 0.02$ & $1.45 \pm 0.01$ & $1.35 \pm 0.01$ & $-1.64 \pm 0.05$ & $-4.03 \pm 0.01 \pm 0.01$ & $0.12 \pm 0.01$ & $0.07 \pm 0.01$ & $28.23 \pm 0.05$ & $4.42 \pm 0.10$ \cr
UGC 03755 & $24.62 \pm 0.12 \pm 0.02$ & $\cdots$ & $\cdots$ & $\cdots$ & $-4.01 \pm 0.03$ & $0.29 \pm 0.03$ & $0.17 \pm 0.02$ & $28.46 \pm 0.13$ & $4.92 \pm 0.29$ \cr
\cutinhead{Estimates made by other authors}
NGC 2366 & $\leq 23.74$ & $\cdots$ & $\cdots$ & $\cdots$ & $-4.01$ & $0.12 \pm 0.01$ & $0.07 \pm 0.01$ & $\leq 27.68 \pm 0.20$ & $\leq 3.44 \pm 0.32$ \cr
UGC 03755 & $24.24$ & $\cdots$ & $\cdots$ & $\cdots$ & $-4.01$ & $0.29 \pm 0.03$ & $0.17 \pm 0.02$ & $28.08 \pm 0.39$ & $4.13 \pm 0.74$ \cr
NGC 2683 & $25.49$ & $\cdots$ & $\cdots$ & $\cdots$ & $-4.01$ & $0.11 \pm 0.01$ & $0.06 \pm 0.01$ & $29.44 \pm 0.36$ & $7.73 \pm 1.28$ \cr
\cutinhead{Lower limits inferred from photometry}
NGC 2366 & $\geq$ 23.00 & $\cdots$ & $\cdots$ & $\cdots$ & $-4.01$ & $0.12 \pm 0.01$ & $0.07 \pm 0.01$ & $\geq$ 26.94 & $\geq$ 2.44 \cr
UGC 03755 & $\geq$ 24.50 & $\cdots$ & $\cdots$ & $\cdots$ & $-4.01$ & $0.29 \pm 0.03$ & $0.17 \pm 0.02$ & $\geq$ 28.34 & $\geq$ 4.65 \cr
NGC 2683 & $\geq$ 25.00 & $\cdots$ & $\cdots$ & $\cdots$ & $-4.01$ & $0.11 \pm 0.01$ & $0.06 \pm 0.01$ & $\geq$ 28.95 & $\geq$ 6.15
\enddata
\label{table:dist}
\end{deluxetable}

\end{document}